\documentclass{aa}
\usepackage{graphicx}
\usepackage[varg]{txfonts}
\begin{document}
\newcommand {\cm} {cm$^{-1}$}
\newcommand {\mc} {$\mu$m}
\newcommand {\wat} {H$_2$O}
\newcommand {\coo} {CO$_2$}
\newcommand {\meoh} {CH$_3$OH}
\newcommand {\met} {CH$_4$}
\newcommand {\oo} {O$_2$}
\newcommand {\nn} {N$_2$}
\newcommand {\nh} {NH$_3$}

   \authorrunning{Ioppolo et al.}
   \titlerunning{VUV spectroscopy of 1~keV electron irradiated CO$_{2}$:H$_{2}$O ices}
   \title{Vacuum ultraviolet photoabsorption spectroscopy of space-related ices: Formation and destruction of solid carbonic acid upon 1~keV electron irradiation}

   %\subtitle{}

   \author{S. Ioppolo  \inst{1}
          \and
          Z.~Ka\v{n}uchov\'{a}  \inst{2}
          \and
          R.L. James  \inst{3}
          \and
          A. Dawes  \inst{3}
          \and
          A. Ryabov  \inst{3}
          \and
          J. Dezalay  \inst{3}
          \and
          N.C. Jones  \inst{4}
          \and
          S.V. Hoffmann  \inst{4}
          \and
          N.J. Mason  \inst{5}
          \and
          G. Strazzulla  \inst{6}
          }
\offprints{S. Ioppolo;  \email{s.ioppolo@qmul.ac.uk}}

\institute{School of Electronic Engineering and Computer Science, Queen Mary University of London, London E1 4NS, UK\\
         \and
Astronomical Institute of Slovak Academy of Sciences, SK-059 60 Tatransk\'{a} Lomnica, Slovakia\\
         \and
School of Physical Sciences, The Open University, Milton Keynes MK7 6AA, UK\\
         \and
ISA, Department of Physics and Astronomy, Aarhus University, Aarhus DK-8000, Denmark\\
         \and
School of Physical Sciences, University of Kent, Canterbury CT2 7NH, UK\\
         \and
INAF - Osservatorio Astrofisico di Catania, Catania I-95123, Italy\\
             }

   %\date{Received September 15, 1996; accepted March 16, 1997}
    \date{Received ; accepted }

% \abstract{}{}{}{}{}
% 5 {} token are mandatory

  \abstract
{Carbonic acid (H$_2$CO$_3$) is a weak acid relevant to astrobiology which, to date, remains undetected in space. Experimental work has shown that the $\beta$-polymorph of H$_2$CO$_3$ forms under space relevant conditions through energetic (UV photon, electron, and cosmic ray) processing of CO$_2$- and H$_2$O-rich ices. Although its $\alpha$-polymorph ice has been recently reassigned to the monomethyl ester of carbonic acid, a different form of H$_2$CO$_3$ ice may exist and is synthesized without irradiation through surface reactions involving CO molecules and OH radicals, that is to say $\gamma$-H$_2$CO$_3$.}
{We aim to provide a systematic set of vacuum ultraviolet (VUV) photoabsorption spectroscopic data of pure carbonic acid that formed and was destroyed under conditions relevant to space in support of its future identification on the surface of icy objects in the Solar System by the upcoming JUpiter ICy moons Explorer (JUICE) mission and on interstellar dust by the James Webb Space Telescope (JWST) spacecraft.}
{We present VUV photoabsorption spectra of pure and mixed CO$_2$ and H$_2$O ices exposed to 1~keV electrons at 20 and 80~K to simulate different interstellar and Solar System environments. Ices were then annealed to obtain a layer of pure H$_{2}$CO$_{3}$ which was further exposed to 1~keV electrons at 20 and 80~K to monitor its destruction pathway. Fourier-transform infrared (FT-IR) spectroscopy was used as a secondary probe providing complementary information on the physicochemical changes within an ice.}
{Our laboratory work shows that the formation of solid H$_{2}$CO$_{3}$, CO, and O$_3$ upon the energetic processing of CO$_{2}$:H$_{2}$O ice mixtures is temperature-dependent in the range between 20 and 80~K. The amorphous to crystalline phase transition of H$_{2}$CO$_{3}$ ice is investigated for the first time in the VUV spectral range by annealing the ice at 200 and 225~K. We have detected two photoabsorption bands at 139 and 200~nm, and we assigned them to $\beta$-H$_2$CO$_3$ and $\gamma$-H$_2$CO$_3$, respectively. We present VUV spectra of the electron irradiation of annealed H$_{2}$CO$_{3}$ ice at different temperatures leading to its decomposition into CO$_{2}$, H$_{2}$O, and CO ice. Laboratory results are compared to Cassini UltraViolet Imaging Spectrograph observations of the $70-90$~K ice surface of Saturn's satellites Enceladus, Dione, and Rhea.}


   \maketitle
%
%________________________________________________________________

\section{Introduction}

Carbonic acid (H$_2$CO$_3$) is a weak acid, which is of great interest in astrochemistry and astrobiology. On Earth, H$_2$CO$_3$ is found in a solution with water (H$_2$O) and carbon dioxide (CO$_2$). It plays an important role in the global carbon cycle and in biological carbonate-containing systems \citep{Wang_etal2016}. For instance, about $30-40\%$ of human-induced emissions of atmospheric CO$_2$ dissolve into the oceans to form carbonic acid \citep{Caldeira_Wickett2003}. Moreover, the chemical weathering of sedimentary rocks mainly occurs due to the action of water, carbonic acid, and oxygen on minerals and rocks \citep{Haldar_Tisljar2014}. In space, carbonic acid is believed to be present in a variety of different environments including the surface of Mercury's north pole, within the Martian polar caps, on the Galilean satellites Europa, Ganymede, and Callisto, on comets, on smaller icy objects in the outer Solar System, and on dust grains in the interstellar medium \citep[ISM, ][]{Strazzulla_etal1996, McCord_etal1997, Hage_etal1998, Zheng_Kaiser2007, Kohl_etal2009, Peeters_etal2010, Jones_etal2014a, Jones_etal2014b, Delitsky_etal2017}. The rationale behind the assumption that carbonic acid is ubiquitous in space is based on extensive laboratory evidence showing that H$_2$CO$_3$ ice is synthesized efficiently upon energetic irradiation of CO$_2$:H$_2$O ice mixtures \citep{Moore_etal1991a, Moore_etal1991b, DelloRusso_etal1993, Brucato_etal1997, Gerakines_etal2000, Wu_etal2003, Garozzo_etal2008, Lv_etal2014}, with water being the most abundant molecule found in astrochemical ices \citep{Boogert_etal2015}.

\cite{Hage_etal1993} produced H$_2$CO$_3$ via acid-base reactions at low temperatures and found two different polymorphs of H$_2$CO$_3$, $\alpha$-H$_2$CO$_3$ in methanolic solutions and $\beta$-H$_2$CO$_3$ in water solutions  \citep{Hage_etal1996a, Hage_etal1996b, Mitterdorfer_etal2012}. $\alpha$-H$_2$CO$_3$ presents weaker absorption bands in the O–H stretch  regions than $\beta$-H$_2$CO$_3$, which has its strongest absorption band around 2600~cm$^{-1}$. However, more recently, the synthesis of $\alpha$-H$_2$CO$_3$ in methanolic solutions was reassigned to the monomethyl ester of carbonic acid \citep[CAME, HO-CO-OCH$_3$, ][]{Reisenauer_etal2014, Kock_etal2020}, leaving the $\beta$-polymorph as the only confirmed form of H$_2$CO$_3$ to date. It is known that H$_2$CO$_3$ synthesized through energetic irradiation of CO$_2$:H$_2$O ice mixtures produces amorphous $\beta$-H$_2$CO$_3$ that crystallises if heated to temperatures above 220~K \citep{Hage_etal1995, Hage_etal1996b, Bernard_etal2013}. \cite{Zheng_Kaiser2007} exposed CO$_2$:H$_2$O ice mixtures to 5~keV electrons in the temperature range $10-60$~K and suggested that $\beta$-H$_2$CO$_3$ forms through a two-step mechanism involving an initial H$_2$O$\cdot$CO$_2$ complex and the formation of the HOCO intermediate

\begin{equation}
		\textrm{H}_2\textrm{O}+\textrm{e}^{-}\rightarrow{\textrm{H}}+\textrm{OH}
	\label{Eq1}
\end{equation}

\begin{equation}
		\textrm{H}+\textrm{CO}_2\rightarrow{cis\textrm{-HOCO}}
	\label{Eq2}
\end{equation}

\begin{equation}
		cis\textrm{-HOCO}\rightarrow{trans\textrm{-HOCO}}
	\label{Eq3}
\end{equation}

\begin{equation}
		trans\textrm{-HOCO}+\textrm{OH}\rightarrow{\textrm{H}_2\textrm{CO}_3.}
	\label{Eq4}
\end{equation}

\cite{Zheng_Kaiser2007} found that the number of H$_2$O$\cdot$CO$_2$ complexes in the mixed ice increases with the deposition temperature and, as a consequence, the formation of H$_2$CO$_3$ follows a similar trend. \cite{Oba_etal2010} proved that H$_2$CO$_3$ can also form without irradiation through the surface reaction of CO molecules and OH radicals at $10-40$~K in a water-rich environment. The structure of H$_2$CO$_3$ obtained in this way differs from that of $\beta$-H$_2$CO$_3$ and it is closer, but not identical, to what was once considered to be the structure of $\alpha$-H$_2$CO$_3$. Hence, a $\gamma$-polymorph might have been produced in the work by \cite{Oba_etal2010} from the reaction of CO ice with OH radicals \citep{Kock_etal2020}

\begin{equation}
		\textrm{CO}+\textrm{OH}\rightarrow{trans\textrm{-HOCO}}
	\label{Eq5}
\end{equation}

\begin{equation}
		trans\textrm{-HOCO}\rightarrow{cis\textrm{-HOCO}}
	\label{Eq6}
\end{equation}

\begin{equation}
		cis\textrm{-HOCO}+\textrm{OH}\rightarrow{\textrm{H}_2\textrm{CO}_3.}
	\label{Eq7}
\end{equation}

According to \cite{Oba_etal2010}, H$_2$CO$_3$ can also be synthesized through surface reaction~\ref{Eq4}, and they found a temperature dependent H$_2$CO$_3$ final yield due to the different mobility of species involved in the temperature range $10-40$~K. In the gas phase, three distinct conformers of H$_2$CO$_3$ exist that differ by the position of the OH hydrogen atoms with the most stable $cis$-$cis$, the slightly less stable $cis$-$trans$, and the least stable $trans$-$trans$ conformers \citep{Bucher_Sander2014}. Energetic processing can inject enough energy in the ice to allow the formation and dimerization of the $cis$-$cis$ conformer leading to the $\beta$-H$_2$CO$_3$ form of ice, even when starting from reaction~\ref{Eq4} \citep{Zheng_Kaiser2007}. On the other hand, the surface reactions~\ref{Eq4}$-$\ref{Eq7} initiated by atom-addition processing, that is without irradiation, at low temperatures can cause the formation of a variety of H$_2$CO$_3$ conformers starting from reactions~\ref{Eq4} and \ref{Eq7} that may stabilize in the ice forming a different structure than the known $\beta$-polymorph \citep{Oba_etal2010, Mitterdorfer_etal2012}. Although $\beta$- and $\gamma$-H$_2$CO$_3$ do not share a common formation pathway and present spectral features that differ in relative intensity and peak position, the existence of $\gamma$-H$_2$CO$_3$ is currently speculative and needs to be confirmed in future work.

Moreover, as previously mentioned, there has been no clear detection of H$_2$CO$_3$ in space. This is likely due to spectral confusion in the mid-infrared (MIR), where the vibrational modes of H$_2$CO$_3$ overlap with those from more abundant species in the solid phase, and the lack of extensive and systematic laboratory data at other spectral frequencies beyond the infrared, such as the vacuum ultraviolet (VUV) spectral range, where H$_2$CO$_3$ can potentially show distinct absorption bands allowing its future identification in the Solar System.

\cite{Jones_etal2014a} reported the ultraviolet-visible (UV-vis, $240-600$~nm) photoabsorption spectrum of solid H$_2$CO$_3$ formed upon 5~keV electron irradiation of a CO$_{2}$:H$_{2}$O~=~5:1 ice mixture at 5.5~K followed by warm up to 216~K. In this case, H$_2$CO$_3$ was unambiguously detected mass spectrometrically in the gas phase upon thermal desorption. However, the UV-vis spectra in the region covered in this study did not present any distinctive signature of H$_2$CO$_3$ ice, but just an increased slope toward shorter wavelengths. More recently, \cite{Pavithraa_etal2019} presented the first VUV spectra of H$_2$CO$_3$ ice in the $120-320$~nm range synthesized by irradiating a CO$_{2}$:H$_{2}$O~=~2:1 ice mixture at 10~K by means of the quasi-monoenergetic light ($\sim$9 eV photons) of the synchrotron source at the National Synchrotron Radiation Centre, Taiwan. The authors identified a photoabsorption band of H$_2$CO$_3$ ice at 200~nm and followed its profile evolution as a function of the sample temperature in the $10-230$~K range. Unfortunately, in this case, H$_2$CO$_3$ photoproduction was conducted exclusively at 10~K. \cite{Zheng_Kaiser2007} pointed-out that temperatures around $10-20$~K do not mimic the majority of Solar System ice surfaces and showed that the H$_2$CO$_3$ formation yield increases with temperature in the range $10-60$~K when CO$_{2}$:H$_{2}$O mixtures are exposed to 5~keV electron irradiation. Moreover, \cite{Jones_etal2014b} showed that 5~keV electron irradiation of a layer of H$_2$CO$_3$ at 80~K leads to its decomposition into CO, CO$_{2}$, and H$_{2}$O ices. Therefore, understanding the formation and destruction pathways of H$_2$CO$_3$ under ISM and Solar System conditions is important if we are to correctly interpret data from the myriad of observational studies expected in the next decade.

Concerning future Solar System missions, the ESA JUpiter ICy moons Explorer (JUICE) mission is set to make detailed observations of Jupiter and three of its largest moons, that are Ganymede, Callisto, and Europa \citep{Banks2012}. Onboard the spacecraft two instruments will cover the VUV-visible range (i.e., the UVS at $55-210$~nm with spectral resolution $\leq0.6$~nm and MAJIS at $400-5400$~nm with spectral resolution $3-7$~nm). Due to the spectral gap of such instruments around $210-400$~nm, it is fundamental to obtain detailed laboratory data covering the full VUV-vis range ($115-750$~nm) to truly support and inform future observations in the Solar System \citep{Hendrix_etal2020}. Moreover, in the MIR, the upcoming NASA James Webb Space Telescope (JWST) mission will map ices with unprecedented spatial and spectral detail, increasing the possibility to detect frozen H$_2$CO$_3$ on water-rich ice grains of the ISM.

Here we present the first systematic study of the formation and destruction of both amorphous and crystalline H$_2$CO$_3$ ice at 20 and 80~K in the VUV spectral range. Complementary MIR data are also provided. We investigate the formation of $\beta$- and $\gamma$-H$_2$CO$_3$ ice under space relevant conditions and discuss their unique fingerprint in the VUV and MIR spectral ranges to support their future detection in space. Keeping in mind that the $\gamma$-polymorph of H$_2$CO$_3$ is yet to be confirmed, hereafter we conveniently name solid H$_2$CO$_3$ synthesized through reactions~\ref{Eq1}$-$\ref{Eq4} as $\beta$-polymorph and H$_2$CO$_3$ ice produced through reactions~\ref{Eq5}$-$\ref{Eq7} as $\gamma$-polymorph.

\section{Experimental}

\subsection{Experimental setup}

The experiments described here were performed using a custom-made portable astrochemistry chamber (PAC), a high vacuum (HV) system with a base pressure of 10$^{-9}$~mbar. The experimental setup is designed to investigate the energetically induced chemistry that takes place in interstellar ices and on the icy surfaces of moons and small objects of the Solar System. More details on the design of the setup can be found in \cite{Ioppolo_etal2020}. In brief, the PAC consists of a compact spherical cube chamber connected to a turbo molecular pump, a closed cycle helium cryostat with a base temperature 20~K and a 1~keV electron gun. At the center of the main chamber, a magnesium fluoride (MgF$_{2}$) or zinc selenide (ZnSe) substrate window is mounted in a holder made of oxygen-free high conductivity copper (OFHC) in thermal contact with the cryostat. The temperature of the substrate is measured with a silicon diode and can be controlled in the range $20-300$~K by means of a Kapton tape heater connected to the OFHC block and regulated with a temperature controller system. The electron beam current of the 1~keV gun was measured at the center of the chamber with a Faraday cup placed instead of the substrate holder, and the flux of electrons used here is $2\times10^{13}$~e$^{-}$/(cm$^{2}$s). A high vacuum pre-chamber, with base pressure $\geq{10}^{-5}$~mbar, is used to prepare gas samples that are then dosed into the main chamber through an all metal leak valve. A rotary stage mounted in between the main chamber and the cryostat head allows for the rotation of the substrate so that it can be positioned with its surface normal to either 1) the inlet tube of the gas deposition line during deposition of gases; 2) the incident light, when a spectrum is acquired; or 3) the electron beam during irradiation.

Experiments were carried-out at two different sites to access both the VUV ($120-340$~nm) and MIR ($4000-600$~cm$^{-1}$) spectral ranges. The advantage of using the same HV system at two different facilities is that experimental conditions and preparation of ices can be accurately reproduced allowing for a more straight forward comparison of results. VUV photoabsorption spectra were measured on the AU-UV beam line at the ASTRID2 synchrotron light source at Aarhus University, Denmark \citep{Eden_etal2006, Palmer_etal2015}. The beam line can produce monochromatized light over the wavelength range 115 to 700~nm using two gratings. The high energy grating settings used for the measurements presented here corresponds to a typical VUV beam flux of 10$^{10}$~photons/s/100~mA and a wavelength resolution of 0.08~nm. The step sizes used for this work were chosen to be between 0.05 and 1~nm, to resolve absorption features of molecules detected in the solid phase. Fourier-transform infrared (FT-IR) spectra of the same ices were measured at the Molecular Astrophysics Laboratory at the Open University (OU), UK \citep{Dawes_etal2016}. A Nicolet Nexus 670 spectrometer coupled to an external mercury cadmium telluride (MCT) detector was used to acquire FT-IR spectra in transmission mode with absorbance spectra collected at 1~cm$^{-1}$ resolution and with an average of 128~scans. Outside the vacuum chamber, IR spectrometer and purge boxes were constantly flushed with compressed dry air to minimize any gas-phase contribution of atmospheric contaminants along the IR-beam line of sight.

\subsection{Experimental procedure}

During a standard experiment, the substrate was first cooled down to 20~K, flash-heated to 200~K, and then cooled to the selected temperature for the experiment to remove possible contaminants from the surface. In this work, experiments were performed at two different temperatures, 20 and 80~K, to simulate ISM and Solar System conditions. Once the substrate was at its selected temperature, a background spectrum was acquired and used as a reference for sample spectra taken afterwards. Pure CO$_2$ and H$_2$O samples, and CO$_2$:H$_2$O  ice mixtures were prepared in a pre-chamber with partial pressures controlled by a mass-independent baratron. Liquid water was deionized and purified (Milli-Q water purifier, EMD Millipore) before being degassed by several freeze-pump-thaw cycles, while gaseous CO$_2$ was used as received from Sigma-Aldrich ($99.995\%$, CANgas). Gases were admitted into the main chamber by an all metal needle valve. During deposition, pressure was controlled and regulated between $(1-5)\times10^{-7}$~mbar by means of a mass-dependent ion gauge, which gave deposition rates of between 0.2 and 1~nm/s.

The thickness of the deposited ice samples at the ASTRID2 facility was determined using a HeNe laser interference technique described elsewhere \cite[e.g.,][]{Born_Wolf1970, Goodman1978, Baratta_Palumbo1998}. The thickness of the ice layer was calculated using the following equation

\begin{equation}
l = {\frac{\lambda_{0}}{2n_{1}\cos\theta_{1}}}\times{N},
	\label{thicknessVUV}
\end{equation}

\noindent
where $\lambda_{0}$ is the wavelength of the Ne-He laser beam  in vacuum (632.8~nm), $\theta_{1}$ is the angle of the laser within the ice, $n_{1}$ the refractive index of the ice film, and $N$ is the number of constructive pattern repetitions during the deposition time. The refractive index $n_{1}$ at 632.8~nm was estimated from the ratio of the maxima and the minima of the laser interference pattern \citep{Born_Wolf1970, Berland_etal1994, Westley_etal1998}. Some of the experimental parameters of the investigated ices, such as the refractive index ($n$), thickness, penetration depth of the electrons in the ice, stopping power (i.e., the energy released by 1~keV electrons into the ice), and the density of the ice, are listed in Table~\ref{T1}. Unfortunately, during the deposition of pure H$_2$O ice at 80~K at ASTRID2, the HeNe laser was not recorded properly by the photodetector, hence, it was not possible to determine the thickness and refractive index of the ice. However, the deposition pressure monitored from inside the vacuum chamber by an ion gauge and the deposition time were kept the same as for the other experiments. Thus, we expect the thickness to be within $2-3$ times that of pure H$_2$O deposited at 20~K (see also Appendix A).

\begin{table*}
\centering
	\caption{Selected parameters of the main experiments.}
	\label{T1}
	\begin{tabular}{lccccccc}
		\hline
        \hline
        Experiment& $n$& \multicolumn{2}{c}{Thickness}&Penetration Depth&\multicolumn{2}{c}{Stopping Power}&Density\\

        && \multicolumn{2}{c}{[\mc]}& [\mc]&[eV/A]& [eV/16u]&[g/cm$^3$]\\

		                                  &    & VUV               & IR                &     &    &10$^{-15}$   &            \\
		\hline
        CO$_2$ at 20 K	                  &1.35&$0.24\pm0.02$$^{a}$&$0.53\pm0.11$$^{b}$&0.04 &2.85&7.79         &0.98$^{c}$  \\
        CO$_2$ at 75 K	                  &1.35&$0.24\pm0.02$$^{a}$&$0.25\pm0.05$$^{b}$&     &    &             &            \\
        H$_2$O at 20 K	                  &1.34&$0.12\pm0.01$$^{a}$&$0.76\pm0.15$$^{b}$&0.06 &1.66&4.45         &1.07$^{d,e}$\\
        H$_2$O at 80 K$^{f}$	          &    &                   &$1.06\pm0.22$$^{b}$&     &	  &             &0.94$^{e,g}$\\
        CO$_2$:H$_2$O~(6:1) at 20 K	      &1.36&$0.96\pm0.09$$^{a}$&$1.87\pm0.38$$^{b}$&0.04 &2.7 &7.37         &1.0$^{h}$   \\
        CO$_2$:H$_2$O~(6:1) at 80 K	      &1.36&$0.96\pm0.09$$^{a}$&$1.57\pm0.32$$^{b}$&     &    &             &1.0$^{h}$   \\
        \hline
                                	      &	   &                   &[nm]               &     &    &10$^{-16}$   &            \\
        \hline
        a-H$_2$CO$_3$ irr at 20 K$^{i,j}$ &	   &                   &$0.79\pm0.16$$^{b}$&0.06 &0.02&0.59         &1.0$^{k}$   \\
        c-H$_2$CO$_3$ irr at 20 K$^{i,j}$ &	   &                   &$4.81\pm0.98$$^{b}$&	 &0.13&3.57         &            \\
        a-H$_2$CO$_3$ irr at 80 K$^{i,j}$ &	   &                   &$1.09\pm0.22$$^{b}$&	 &0.03&0.81         &            \\
        c-H$_2$CO$_3$ irr at 80 K$^{i,j}$ &	   &                   &$2.41\pm0.49$$^{b}$&     &0.07&1.79         &            \\
		\hline
	\end{tabular}
\begin{flushleft}
\footnotetext{}{$^{a}$The uncertainty in the ice thickness values is calculated by assuming a 10\% error in the estimation of $n$; $^{b}$the uncertainty in the ice thickness values is calculated by assuming a 10\% error in the band strength and density of the ice; $^{c}$\cite{Luna_etal2012}; $^{d}$\cite{Narten_etal1976}; $^{e}$\cite{Jenniskens_Blake1994}; $^{f}$laser measurement not recorded; $^{g}$\cite{Sceats_Rice1982}; $^{h}$average value for a CO$_2$:H$_2$O~(6:1) mixture; $^{i}$a- and c-H$_2$CO$_3$ stand for amorphous and crystalline H$_2$CO$_3$ ice, synthesized at either 20 or 80~K and then annealed to 200 and 220~K, respectively; $^{j}$solid H$_2$CO$_3$ is thinner than the penetration depth of 1~keV electrons; $^{k}$assumed value \citep{Peeters_etal2010}.}
\end{flushleft}
\end{table*}

The ice thickness of samples deposited at the OU was determined by calculating first the column densities of the ice species from the MIR spectra by integrating the Beer-Lambert law equation over the absorption band for a given vibrational transition in the following rearranged equation

\begin{equation}
N_i = \frac{1}{A_i}\int\tau(\nu)\: \mathrm{d}\nu,
  \label{columndensity}
\end{equation}

\noindent
where $N_i$ is the column density [molecules/cm$^{2}$], the integral $\int\tau(\nu)~\:\mathrm{d}\nu$ corresponds to the area of the absorption band for a given vibrational mode of a molecular species, $\tau(\nu)$ is the optical depth of the material, \textit{i}, and $A_i~=~\int\sigma(\nu)\:~\mathrm{d}\nu$ is the band strength for the selected absorption band obtained from the literature. The thickness of an ice was then calculated by scaling the column density by a conversion factor ($N_V$) defined as the number density per unit of volume

\begin{equation}
  l = \frac{N_i}{N_V} = {N_i}\times{\frac{Z m_p}{\rho_i}},
  \label{thicknessIR}
\end{equation}

\noindent
where $Z$ is the atomic number [amu], $m_p$ is the proton mass [g], and $\rho_i$ is the density of the ice [g/cm$^3$]. From the column density of the ice species in the MIR it was also possible to determine the ratios of the ice mixtures studied.

\begin{figure}
\centering
%\hspace{-10.5mm}
\includegraphics[width=0.5\textwidth]{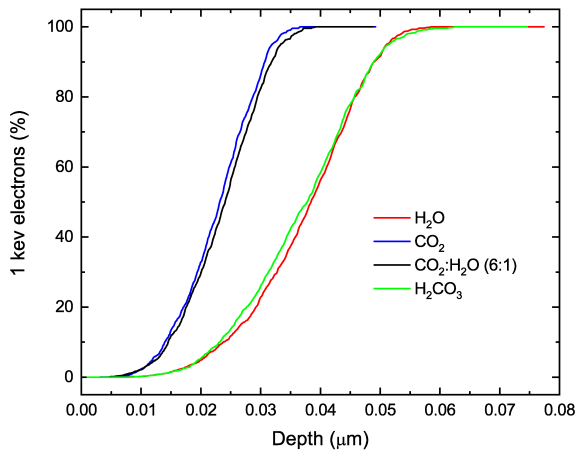}
\vspace{-5mm}
\caption{Results from CASINO simulations of 1~keV electrons trajectories in pure CO$_2$ and H$_2$O ices, in a CO$_2$:H$_2$O~(6:1) ice mixture and in H$_2$CO$_3$ ice. For each simulation 1000 trajectories were calculated. All electrons impinging normal to the pure CO$_2$ and CO$_2$:H$_2$O~(6:1) ice surfaces are stopped at 0.04~\mc\ (solid blue and black lines, respectively), while all electrons are stopped in pure H$_2$O and H$_2$CO$_3$ ices at 0.06~\mc\  (solid red and green lines, respectively).}
\label{F1}
\end{figure}

After deposition, an absorbance spectrum of the deposited sample was acquired. Samples were then exposed to the electron beam of a 1~keV electron gun for three hours for a total fluence of $2.2\times10^{17}$~e$^{-}$/cm$^{2}$ corresponding to a dose of $9.9\times10^{2}$~eV/16u. The penetration depth of 1~keV electrons was estimated using the CASINO software (Monte Carlo simulation of electron trajectory in solids) to be around 0.04~\mc\ for both pure solid CO$_2$ and CO$_2$:H$_2$O~(6:1) ice mixtures, and 0.06~\mc\ for both pure H$_2$O and H$_2$CO$_3$ ices (see Figure~\ref{F1} and Table~\ref{T1}). Since the ice thicknesses of pure and mixed ices were larger than the penetration depth of the impinging electrons, the electrons were implanted into the ice upon irradiation. Absorption spectra were acquired at the end of the irradiation exposure. The irradiated ices were subsequently heated to either 200 or 225~K and then cooled down to the selected temperature, that were 20 or 80~K. The annealing process was used to allow CO$_{2}$ and H$_{2}$O to desorb from the sample surface leaving the most refractory irradiation products, that was frozen H$_{2}$CO$_{3}$, in the solid phase \citep{Gerakines_etal2000}.

We have measured the area of the 2626~cm$^{-1}$ absorption band of solid H$_{2}$CO$_{3}$ and used its band strength, that is $16\times10^{-17}$~cm/molecule \citep{Gerakines_etal2000}, to calculate the column density of the synthesized H$_{2}$CO$_{3}$ using a rearranged form of the Beer–Lambert Law, that is equation~\ref{columndensity}. The thickness of the H$_{2}$CO$_{3}$ layers was estimated to be in the $0.5-2$~nm range, with an assumed density of 1~g/cm$^3$ (see Table~\ref{T1}). A spectrum of H$_{2}$CO$_{3}$ at 20 or 80~K was then used as a reference for sample spectra taken afterwards. At this point, the ice was exposed to electron irradiation again for a time between 30 minutes and 3 hours, depending on the specific experiment, which corresponds to fluences between $3.6\times10^{16}$ and $2.2\times10^{17}$~e$^{-}$/cm$^{2}$. The second irradiation was performed to determine products formed by electron irradiation of H$_{2}$CO$_{3}$ ice \citep{Jones_etal2014b}. It is worthy of note that most of the effects produced by energetic ions interacting with solid matter are due to secondary electrons produced along the ion track in the irradiated sample \citep{Baratta_etal2002}. Thus, studying the effect of electron exposure to ice samples is an essential step towards a better understanding of a variety of energetic processing occurring in space.

\section{Results and discussion}

Since CO$_2$ and H$_2$O ice mixtures are ubiquitously observed in space, a detailed study of their interaction with energetic particles, UV photons, cosmic rays, and electrons, is important to our understanding of chemistry in space \citep[e.g., ][]{Gibb_etal2004, Pontoppidan_etal2008, Ioppolo_etal2013}. Here we investigate the electron irradiation of pure and mixed CO$_2$ and H$_2$O ices to discuss the formation and destruction of frozen carbonic acid and provide the first systematic set of VUV data for the future observation of this species in space.

\subsection{Pure CO$_2$ and H$_2$O ices}

\begin{figure}
\centering
%\hspace{-10.5mm}
\includegraphics[width=0.5\textwidth]{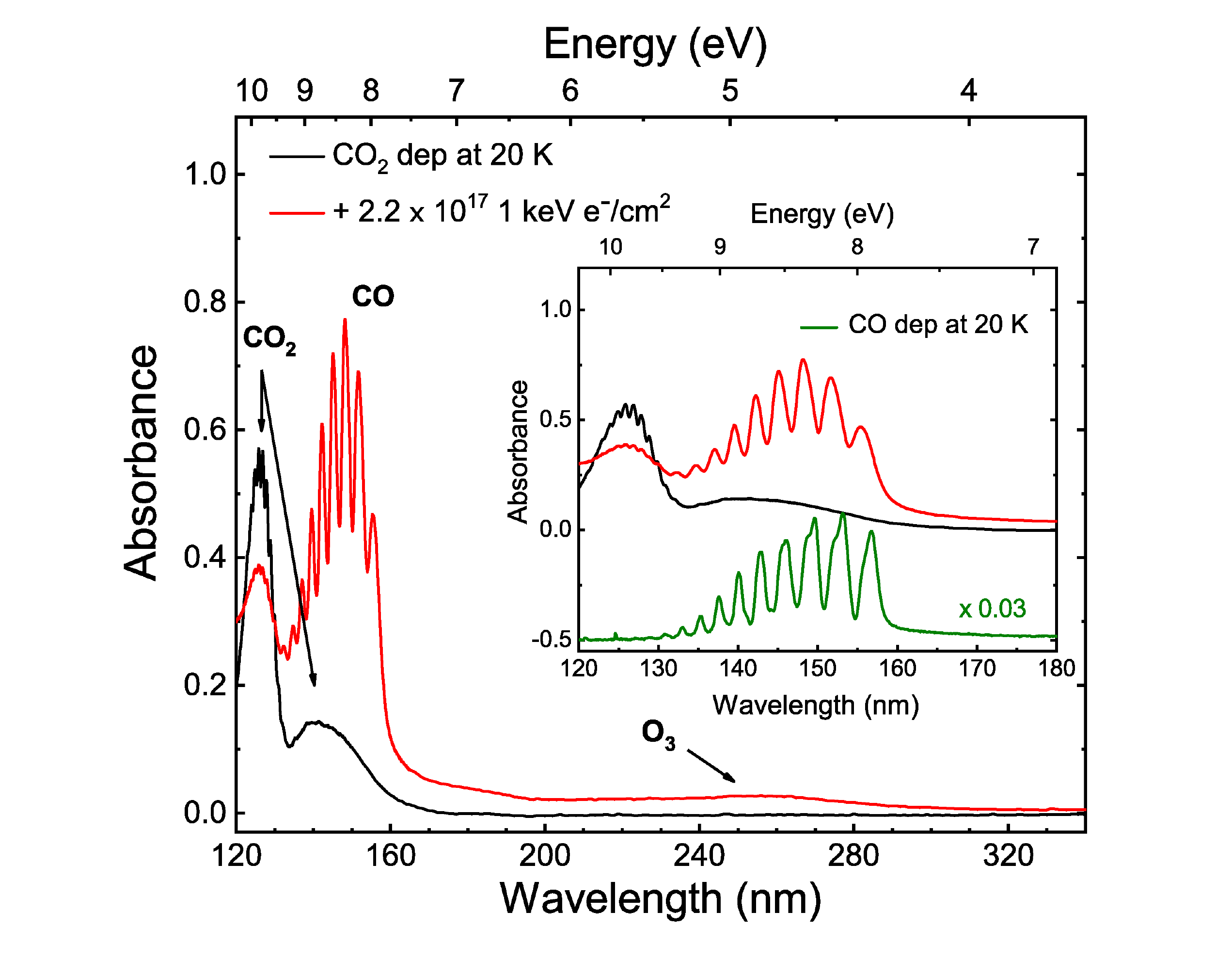}
\vspace{-5mm}
\caption{VUV ($120-340$~nm) spectra of pure CO$_2$ ices deposited at 20~K before and after (solid black and red lines, respectively) 1~keV electron irradiation with a total fluence of $2.2\times10^{17}$~e$^{-}$/cm$^{2}$, i.e., a dose of $9.9\times10^{2}$~eV/16u. A step size interval of 0.05~nm is used to resolve the fine structures in the spectral range between $120-160$~nm. The reminder of the VUV spectra are acquired with a step size of 1~nm. A VUV spectrum of pure CO ice acquired with a step size of 0.1~nm is also displayed for comparison (solid green line).}
\label{F2}
\end{figure}

Figure~\ref{F2} shows a VUV photoabsorption spectrum of solid CO$_2$ deposited at 20~K (solid black line) in the spectral range $120-340$~nm with an inset showing the spectrum of pure CO$_2$ in the $120-180$~nm range. \cite{Mason_etal2006} showed that CO$_2$ ice presents two broad absorption features in the VUV. Centered at 126 and 140~nm, such spectral bands are assigned to the \textrm{${^{1}\Pi_{\textrm{g}}}\leftarrow{^{1}\Sigma^{+}_{\textrm{g}}}$} transition due to the promotion of an electron from the \textrm{$1\pi_\textrm{g}$} to the \textrm{$3\textrm{s}\sigma_\textrm{g}$} orbitals and to the ${^{1}\Delta_{\textrm{u}}}\leftarrow{^{1}\Sigma^{+}_{\textrm{g}}}$ transition due to the promotion of a \textrm{$1\pi_\textrm{g}$} electron to the \textrm{$2\pi_\textrm{u}$} orbital \citep{Yoshino_etal1996}. Although the broad CO$_2$ transition at 140~nm appears quite smooth, the higher energy transition around 126~nm presents extensive vibrational structure \citep{Lu_etal2008}.

Upon 1~keV electron irradiation, the 126~nm band decreases in intensity and a new strong vibrational progression appears in the same spectral region of the broad CO$_2$ band at 140~nm (solid red line). A comparison with a VUV spectrum of pure CO ice deposited at 20~K (solid green line) reveals that this progression is due to newly synthesized CO molecules in a CO$_2$ ice and can be assigned to the ${\widetilde{\textrm{A}}^{1}\Pi}\leftarrow{\widetilde{\textrm{X}}^{1}\Sigma^{+}}$ transition \citep[see Table~\ref{T2} for peak positions, ][]{Lu_etal2005, Mason_etal2006, CruzDiaz_etal2014a}. The VUV vibrational progression of CO ice is recorded with a step size interval of 0.05~nm to highlight its fine structure. The VUV spectrum of photoprocessed CO$_2$ ice deposited at 8~K as presented by \cite{CruzDiaz_etal2014b} is qualitatively similar to our spectrum of electron irradiated ice, confirming formation of CO in the ice. As discussed in \cite{CruzDiaz_etal2014b}, there is a shift to shorter wavelengths in the vibrational progression of CO mixed with CO$_2$ with respect to the pure CO transitions (see inset of Figure~\ref{F2} and Table~\ref{T2}). Moreover, the relative intensity of the bands changes. Another product of the electron irradiation of pure CO$_2$ is frozen ozone (O$_3$) identified through the absorption feature at 257~nm corresponding to the Hartley band due to the ${1^{1}\textrm{B}_{2}}\leftarrow{\textrm{X}^{1}\textrm{A}_{1}}$ transition \citep[see Figure~\ref{F2}, ][]{Mason_etal1996}.

The presence of CO and O$_3$ in the ice suggests that CO$_2$ is dissociated by 1~keV electrons into CO molecules and O atoms. A detailed discussion on the radiolysis-induced dissociation reactions of CO$_2$-containing ices is presented in \cite{Bennett_etal2010}. The latter can then recombine forming O$_2$ and O$_3$ molecules. At 20~K, the newly formed CO, O$_2$, and O$_3$ do not desorb and should all be detectable in the VUV spectral range. However, O$_2$ ice has a strong absorption band at 150~nm that overlaps with the vibrational progression of CO, making its unambiguous detection harder \citep{Ioppolo_etal2020}. At 75~K, only traces of CO$_2$ molecules and other species are detected in the ice during 1~keV electron processing of the ice (see Figure~\ref{S1}). This means that the electron-induced dissociation products desorb efficiently upon formation due to the higher temperature of the ice \citep{Collings_etal2004, Acharyya_etal2007, Noble_etal2012, Jones_etal2014a, Collings_etal2015}.

\begin{table}
\centering
	\caption{CO ice electronic transitions observed in the VUV photoabsorption spectra of CO$_2$ ice deposited at 20 and 75~K and then irradiated with 1~keV electrons at a total dose of $9.9\times10^{2}$~eV/16u in the $120-160$~nm spectral range. Data are compared to the transition of pure CO ice deposited at 20~K.}
	\label{T2}
	\begin{tabular}{llllll}
		\hline
        \hline
        \multicolumn{2}{c}{Pure CO}& \multicolumn{2}{c}{CO$_2$ irr., 20~K}& \multicolumn{2}{c}{CO$_2$ irr., 75~K}\\

		[nm] & [eV] & [nm] & [eV] & [nm] & [eV]\\
		\hline
		127.1&  9.75&      &      &      &     \\
		129.0&  9.61&	   &      &      &     \\
        130.8&  9.48&	   &      &      &     \\
        133.0&  9.32& 132.4&  9.36&      &     \\
        135.3&  9.16& 134.8&  9.20&      &     \\
        137.6&  9.01& 137.1&  9.04&	138.1& 8.98\\
        140.1&  8.85& 139.6&  8.88&	140.4& 8.83\\
        142.8&  8.68& 142.3&  8.71&	143.1& 8.66\\
        145.6&  8.52&	   &      &      &     \\
        146.1&  8.49& 145.1&  8.54&	145.7& 8.51\\
        148.6&  8.34&	   &      &      &     \\
        149.6&  8.29& 148.2&  8.37&	148.9& 8.33\\
        152.0&  8.16&	   &      &      &     \\
        153.1&  8.10& 151.7&  8.17&	152.0& 8.16\\
        156.7&  7.91& 155.5&  7.97&	154.9& 8.00\\
		\hline
	\end{tabular}
\end{table}

\begin{figure}
\centering
%\hspace{-10.5mm}
\includegraphics[width=0.5\textwidth]{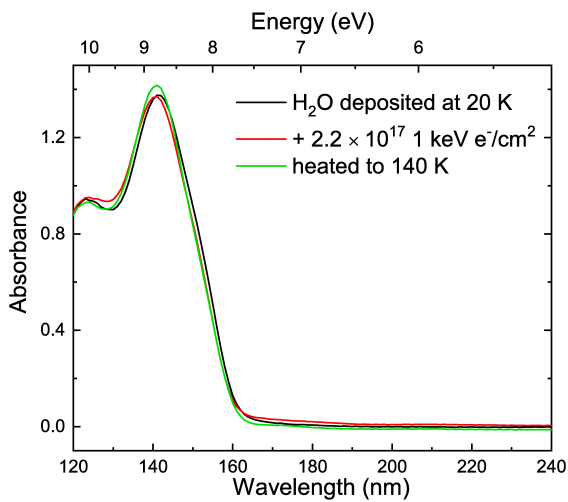}
\vspace{-5mm}
\caption{VUV ($120-240$~nm) spectra of pure H$_2$O ices deposited at 20~K before and after (solid black and red lines, respectively) 1~keV electron irradiation with a maximum total fluence of $2.2\times10^{17}$~e$^{-}$/cm$^{2}$. A VUV photoabsorption spectrum of the processed ice heated to 140~K is also shown for comparison (solid green line). All VUV spectra are acquired with a step size of 1~nm.}
\label{F3}
\end{figure}

Figure~\ref{F3} presents VUV photoabsorption spectra of water ice. As in the case of pure CO$_2$ ice, amorphous solid water (ASW) is deposited at 20~K and then further exposed to a 1~keV electron beam with a total dose of $9.9\times10^{2}$~eV/16u (solid black and red lines, respectively). The ice is then annealed to higher temperatures until complete desorption occurred. The VUV photoabsorption spectrum of high-density ASW at 20~K presents a band in the $120-132$~nm range attributed to the transition ${\widetilde{\textrm{B}}^{1}\textrm{A}_{1}}\leftarrow{\widetilde{\textrm{X}}^{1}\textrm{A}_{1}}$ \citep{Lu_etal2008}. At 132~nm, there is a local minimum, which is also observed by \cite{Lu_etal2008}. The strongest absorption band of ASW in the VUV at 144~nm is assigned to the ${\widetilde{\textrm{A}}^{1}\textrm{B}_{1}}\leftarrow{\widetilde{\textrm{X}}^{1}\textrm{A}_{1}}$ \citep{Mota_etal2005, Mason_etal2006, Lu_etal2008, CruzDiaz_etal2014a}.

\cite{Palumbo2006} showed the formation of compact solid water after ion irradiation of porous ASW at 15~K. Starting from cubic crystalline ice (I$_\textrm{c}$), \cite{Baratta_etal1991} and \cite{Leto_Baratta2003} proved that both ion and UV-photon irradiation induce amorphization of I$_\textrm{c}$ when irradiated at lower temperatures. Although we do not exclude morphological changes in the ice upon electron irradiation, under our experimental conditions and settings, there is no evidence for major changes in the spectral profiles of the VUV photoabsorption water bands before and after 1~keV electron irradiation. Hence, H atoms and OH radicals, that are the electron-induced dissociation products of pure ASW ice, recombine efficiently at 20~K. At 80~K, the decrease in intensity of the 144~nm water band upon electron exposure of the ice indicates that some desorption occurs (see Figure~\ref{S1}). Both H and OH are volatile species that if synthesized at the surface of the ice can recombine or sublimate \citep[see Figure~\ref{S1}, ][]{Andersson_vanDishoeck2008}. Some spectral changes are seen upon heating the ice before its full desorption. Particularly, the VUV spectrum of ASW electron irradiated at 20~K and heated to 140~K (solid green line in Figure~\ref{F3}) presents a slightly sharper feature at 144~nm. Such spectral modifications are likely due to a phase change in the ice at higher temperatures because an amorphous solid water ice layer annealed to 140~K undergoes restructuring toward crystallization before desorption \citep{Jenniskens_Blake1994, Fraser_etal2001, Dohnalek_etal2003, Allodi_etal2014}. \cite{Zheng_etal2006a, Zheng_etal2006b} investigated the electron irradiation of cubic crystalline water ice under ultrahigh vacuum conditions at temperatures between 12 and 90~K and observed a minor production of hydrogen and oxygen, both molecular and atomic, and of hydrogen peroxide that decreases with increasing temperature. These findings strongly indicate a thermal, possibly diffusion-controlled component of the reaction mechanism, which could facilitate the back-reaction toward the reformation of water ice in agreement with our results.

\subsection{Mixed CO$_2$:H$_2$O ices}

\begin{figure}
\centering
%\hspace{-10.5mm}
\includegraphics[width=0.5\textwidth]{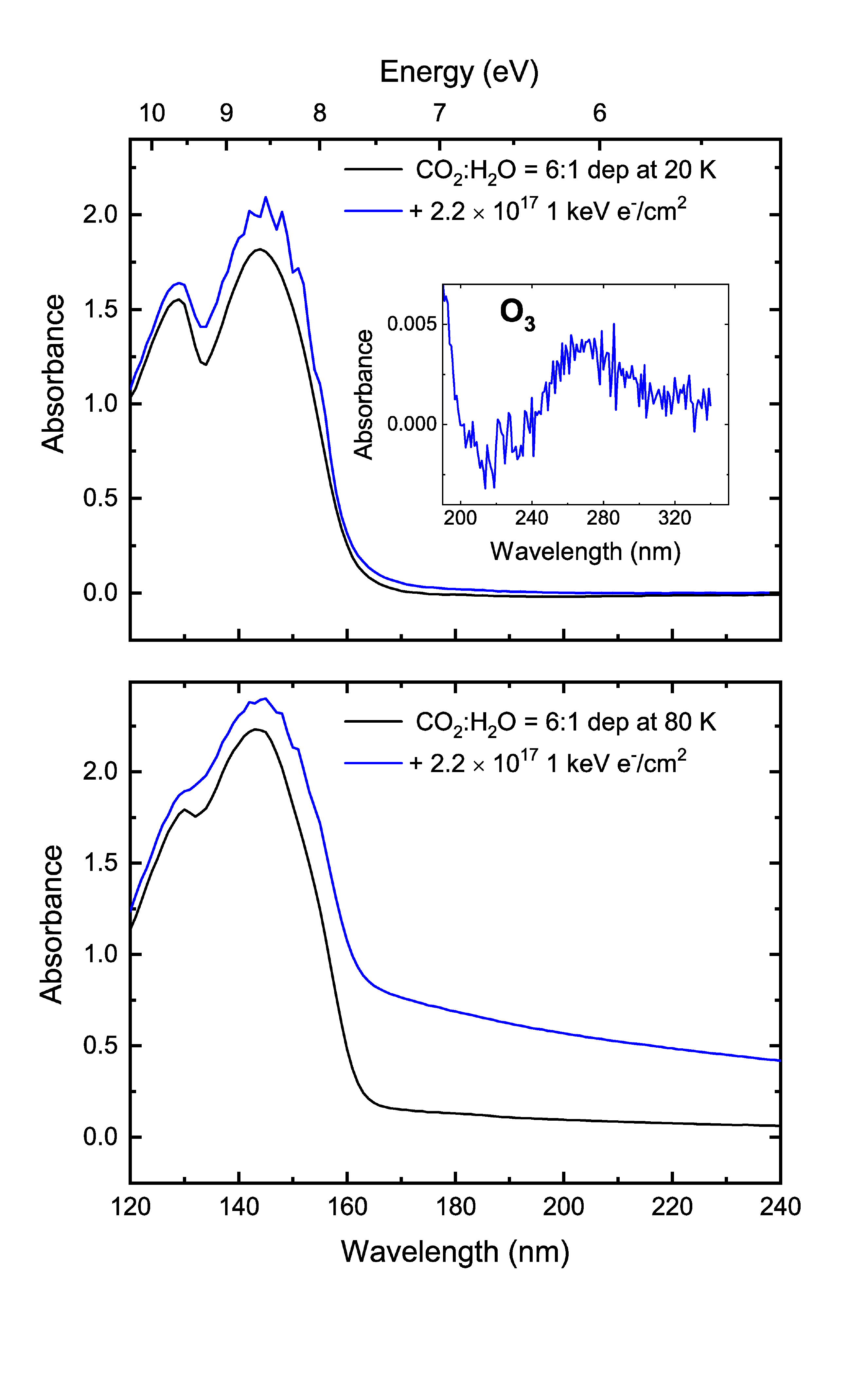}
\vspace{-5mm}
\caption{VUV ($120-240$~nm) photoabsorption spectra of CO$_2$:H$_2$O~(6:1) ice mixtures deposited at 20 and 80~K (top and bottom panels, respectively) before and after (solid black and blue lines respectively) 1~keV electron irradiation with a total fluence of $2.2\times10^{17}$~e$^{-}$/cm$^{2}$. All VUV spectra are acquired with a step size of 1~nm.}
\label{F4}
\end{figure}

To achieve the goal of this work that is the systematical study of the formation and destruction of solid H$_{2}$CO$_{3}$ in the VUV spectral range, we have first performed an extensive list of control experiments on CO$_{2}$:H$_{2}$O ice mixtures with different ratios (1:1, 3:1, 5:1, 6:1, 10:1, and 20:1) irradiated with 1~keV electrons for periods between 1 and 6 hours corresponding to doses of $(3.3-19.8)\times10^{2}$~eV/16u, and subsequently annealed to 220~K. The control experiments allowed us to identify the experimental conditions that maximize the production of H$_{2}$CO$_{3}$ ice upon 1~keV electron irradiation under our experimental settings, that are an initial CO$_{2}$:H$_{2}$O~=~6:1 ice mixture and a total dose of $9.9\times10^{2}$~eV/16u. Although, in the ISM, water is the most abundant molecule in the solid phase, our choice of irradiating CO$_2$:H$_2$O~(6:1) ice mixtures at different temperatures is still astronomically relevant. In many astrophysical environments, as for example icy satellites and other minor objects in the outer Solar System or ice mantles on interstellar grains, solid molecules undergo heating, which causes the segregation of volatile species into more refractory ices. Thus, patches of CO$_2$-rich ice are expected to exist on the surface of the above mentioned objects.

The top panel of Figure~\ref{F4} presents the VUV photoabsorption spectra of a CO$_2$:H$_2$O~(6:1) ice mixture deposited at 20~K before and after 1~keV electron irradiation (solid black and blue lines, respectively). The step size interval of the VUV photoabsorption spectra is 1~nm. Therefore, as opposed to Figure~\ref{F2}, in Figure~\ref{F4} the vibrational progression of solid CO formed upon electron exposure of the ice is not fully resolved. Moreover, the VUV spectra of all the deposited ice mixtures are quite thick (i.e., around 1~$\mu$m, see Table~\ref{T1}). Therefore, some of the feature profiles are potentially affected by saturation effects. We note that the electron irradiation fluence is the same as for the VUV experiments of pure CO$_2$ and H$_2$O ices. As for the experiment of electron bombardment of pure CO$_2$ ice, the VUV spectrum after irradiation of the ice mixture at 20~K reveals the formation of solid CO and O$_3$ (see top panel and its inset in Figure~\ref{F4}). Moreover, the VUV spectrum of the deposited mixture at 80~K is not much different from the one deposited at 20~K and the VUV spectrum of the electron irradiated mixture at 80~K confirms what has been seen in the case of pure CO$_2$ ice at 75~K, that is the formation of CO and its sublimation at 80~K. Traces of CO molecules trapped in the ice mixture are visible in the bottom panel of Figure~\ref{F4}, see the weaker CO vibrational progression. At 80~K, ozone is not detected in the VUV spectra. Hence, it is likely not synthesized in the ice because of the thermal desorption of its parent species, that are O and O$_2$, upon formation. Finally, a clear slope at wavelengths higher than 160~nm appears in the VUV spectrum of the irradiated ice at 80~K. This slope is most likely to be caused by Rayleigh scattering of light from a rougher ice surface obtained after irradiation at 80~K and is not present in the VUV spectrum at 20~K.

\begin{figure}
\centering
%\hspace{-10.5mm}
\includegraphics[width=0.5\textwidth]{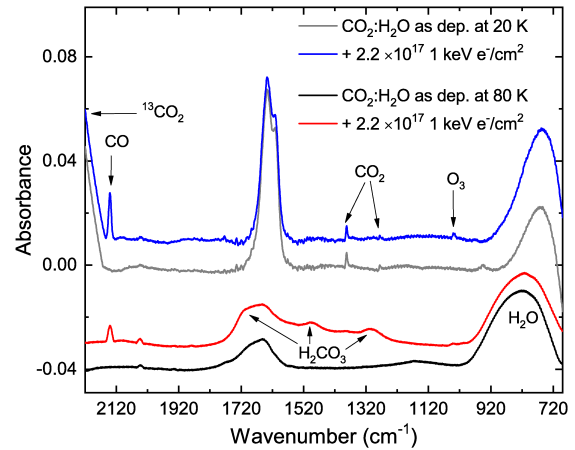}
\vspace{-5mm}
\caption{MIR ($2220-690$~cm$^{-1}$) FT-IR spectra of CO$_{2}$:H$_{2}$O~(6:1) ice mixtures deposited at 20 and 80~K before (solid gray at 20~K and black lines at 80~K) and after (solid blue at 20~K and red lines at 80~K) 1~keV electron irradiation with a total fluence of $2.2\times10^{17}$~e$^{-}$/cm$^{2}$. The inset shows the difference spectra of the CO$_{2}$:H$_{2}$O~(6:1) ice at 20~K after and before electron exposure.}
\label{F5}
\end{figure}

In Figure~\ref{F5}, MIR spectra of the CO$_2$:H$_2$O~(6:1) ice mixture deposited at 20~K before and after 1~keV electron irradiation (solid gray and blue lines, respectively) are shown in the $2220-690$~cm$^{-1}$ range. In this region, the water bending mode of the deposited ice at 1630~cm$^{-1}$ appears split in two peaks, one being due to a H$_2$O$\cdot$CO$_2$ complex at 1637~cm$^{-1}$ and the other due to pure H$_2$O ice at 1610~cm$^{-1}$. Here the CO$_2$ symmetric stretch (${\nu_1}$) is also visible at 1382~cm$^{-1}$ because CO$_2$ is mixed with water ice and the sample is thick enough to see this weak feature that otherwise in pure CO$_2$ ice is infrared inactive \citep{McGuire_etal2016}. There is another H$_2$O$\cdot$CO$_2$ complex at 1275~cm$^{-1}$ and the libration mode of water ice is observed at 755~cm$^{-1}$ \citep{Zheng_Kaiser2007}. The electron irradiated ice MIR spectrum shows the formation of CO and O$_3$ from their stretching modes at 2140 and 1040~cm$^{-1}$, respectively. The presence of H$_2$CO$_3$ in the ice is detected through the observation of several absorption bands in the inset of Figure~\ref{F5}, where the difference spectra between the FT-IR spectrum after and before electron exposure is shown. H$_2$CO$_3$ present several absorption bands at 1723, 1486, and 1292~cm$^{-1}$ shown in the inset of Figure~\ref{F5}, that are the C$=$O stretch, the C$-$OH asymmetric stretch and the C$-$OH in-plane bend, respectively \citep{Gerakines_etal2000, Zheng_Kaiser2007}. Although the 1723~cm$^{-1}$ band is usually the strongest of the series, this is not the case in our difference spectra because of changes in the more intense H$_2$O and H$_2$O$\cdot$CO$_2$ complex absorption bands overlapping with H$_2$CO$_3$. The inset of Figure~\ref{F5} shows also the tentative detection of $cis$- and $trans$-HOCO at 1773 and 1848~cm$^{-1}$, respectively, with the latter being the less intense \citep{Milligan_Jacox1971}. The presence of both isomers of the HOCO radical suggests that solid H$_2$CO$_3$ is formed through both reactions~\ref{Eq4} and \ref{Eq7} under our experimental conditions at 20~K. In this scenario, H$_2$CO$_3$ forms through the H$_2$O$\cdot$CO$_2$ complex route and also when thermalized OH radicals react barrierlessly with the frozen CO, previously synthesized in the ice at 20 K. The different conformers of solid H$_2$CO$_3$ formed throught these routes can stabilize in the ice at 20 K, especially because they are likely to be synthesized in isolation within a CO$_2$:H$_2$O ice matrix.

At 80~K, as expected, none of the isomers of HOCO are detected upon electron exposure of the ice mixture because of the higher temperature of the ice that induces diffusion and desorption of volatile species (solid red line in Figure~\ref{F5}). However, three absorption bands due to H$_2$CO$_3$ are clearly visible at 1710, 1495, and 1307~cm$^{-1}$. By comparing the spectra before and after irradiation at 80~K (solid black and red lines), it becomes clear that CO is formed in the ice and a fraction of it stays trapped in the ice matrix at 80~K. A negligible amount of solid O$_3$ is visible. Differences in CO and O$_3$ formation yields between pure and mixed ices compared at the same temperatures can be explained by the presence of water in the ice mixture that favors trapping and reaction of parent and product species.

\subsection{Formation of pure H$_2$CO$_3$ ice}

\begin{figure}
\centering
%\hspace{-10.5mm}
\includegraphics[width=0.5\textwidth]{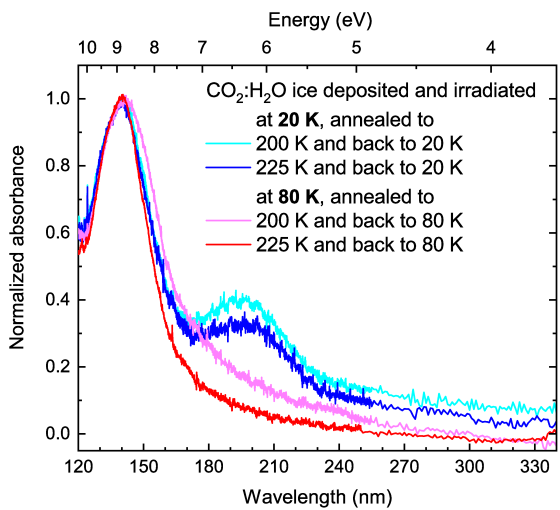}
\vspace{-5mm}
\caption{VUV ($120-340$~nm) photoabsorption spectra of pure H$_2$CO$_3$ ices synthesized upon 1~keV electron irradiation of CO$_2$:H$_2$O~(6:1) ice mixtures at 20 and 80~K and subsequently annealed to 200 and 225~K to obtain amorphous and crystalline ice, respectively. All VUV spectra are normalized to the 139~nm peak of solid H$_2$CO$_3$. VUV spectra are acquired with a step size interval of 0.05~nm between $120-250$~nm and an interval of 1~nm between $250-340$~nm.}
\label{F6}
\end{figure}

Since H$_2$CO$_3$ is more refractory than CO$_2$ and H$_2$O, an effective way of producing a pure sample of H$_2$CO$_3$ is to heat the ice until full desorption of water and carbon dioxide \citep{Moore_etal1991a}. Figure~\ref{F6} shows the VUV absorption spectra of solid pure H$_2$CO$_3$ obtained after annealing the processed CO$_2$:H$_2$O~(6:1) ice mixtures discussed in the previous section. The figure compares four different H$_2$CO$_3$ formation experiments, where ice mixtures are deposited and electron processed at 20 and 80~K and then annealed to 200 and 225~K. The choice of annealing the ice to two different temperatures is made to study different morphological structures of solid H$_2$CO$_3$ in the VUV spectral range \citep[amorphous versus crystalline, ][]{Bernard_etal2013}. All spectra shown in Figure~\ref{F6} present an absorption peak at 139~nm, which we assign to H$_2$CO$_3$ ice. Here all solid H$_2$CO$_3$ spectra are normalized to this peak. To date, the only VUV data on pure H$_2$CO$_3$ ice available in literature are reported by \cite{Pavithraa_etal2019}, who did not detect the 139~nm feature in their VUV spectra. Figure~2 of their paper shows several H$_2$CO$_3$ VUV spectra during heating. Looking more closely at the $130-140$~nm region, a shoulder of an absorption feature is actually visible. However, any data points below 130 nm seem to be very noisy, likely because of the lower amount of VUV photons reaching the detector at $\leq$130~nm in \cite{Pavithraa_etal2019}.

Interestingly, in Figure~\ref{F6} there is a second absorption feature at 200~nm only in the spectra of ices deposited and processed at 20~K, regardless of the final annealing temperature. \cite{Pavithraa_etal2019} identified the 200~nm peak as frozen H$_2$CO$_3$. We agree with their assignment and point out that this feature disappears when H$_2$CO$_3$ is synthesized at higher temperatures. \cite{Pavithraa_etal2019} performed their experiments only at 10~K, therefore, they could not have noticed such behavior. Their data is, however, in good agreement with our VUV spectra of H$_2$CO$_3$ synthesized at 20~K. To better understand this on/off effect of the 200~nm photoabsorption band of pure H$_2$CO$_3$ at different temperatures, we need to consider the formation mechanisms involved and the composition of the ice at such temperatures. As confirmed by our data, formation of H$_2$CO$_3$ at 80~K most likely occurs through the mechanism suggested by \cite{Zheng_Kaiser2007}, that is a two-step mechanism involving an initial H$_2$O$\cdot$CO$_2$ complex and reactions~\ref{Eq1}$-$\ref{Eq4}. Such process leads to the formation of $\beta$-H$_2$CO$_3$ \citep{Hage_etal1996b}. This is also confirmed in our experiments by the MIR peak position of the H$_2$CO$_3$ absorption bands shown in Figure~\ref{F5}. However, at $10-20$~K, solid H$_2$CO$_3$ can also be formed through reactions~\ref{Eq5}$-$\ref{Eq7} involving newly formed CO ice that does not desorb from the ice at temperatures below 30~K \citep{Acharyya_etal2007}. As pointed out by \cite{Oba_etal2010}, this reaction pathway leads to the formation of a different form of H$_2$CO$_3$, namely the $\gamma$-polymorph \citep{Kock_etal2020}. The H$_2$CO$_3$ synthesized at 20~K in our experiments is likely a combination of $\beta$- and $\gamma$-polymorph ice. Hence, we assign the 139~nm band to $\beta$-H$_2$CO$_3$ ice and the 200~nm band to frozen $\gamma$-H$_2$CO$_3$. This assignment is further confirmed by a systematic comparison of FT-IR data of H$_2$CO$_3$ ices discussed later in the text. We exclude any of the 139 and 200~nm bands to be due to H$_2$CO$_3$ hydrates, because MIR data shows that the amount of water trapped in the ice is not linked to the H$_2$CO$_3$ formation temperature, since, for instance, water is also trapped during the experiments at 80~K (see the section on the destruction of pure H$_2$CO$_3$). Moreover, the 200~nm band should not be due to the H$_2$CO$_3$ monomer because the absorption band does not disappear during annealing of the ice above 220~K. The band is indeed present in both amorphous and crystalline H$_2$CO$_3$ ice and it is independent on the crystallization process. We have repeated the H$_2$CO$_3$ formation experiments monitoring the ice in the VUV and confirming the presence of the 200~nm band only in electron irradiation experiments carried out at 20~K.

Concerning phase transition between amorphous and crystalline H$_2$CO$_3$ ice, VUV spectra of the analog experiments annealed to 200 and 225~K do not show any major differences. In the processed ices at 20~K, the one annealed at 225 K presents a relatively slightly weaker 200 nm band compared to the one from the ice annealed at 220~K. In the case of experiments carried out at 80~K, the ice annealed at 225~K shows a sharper 139~nm band and less of a tail at lower wavelengths. Clear differences between amorphous and crystalline H$_2$CO$_3$ ice will be presented later in the text, when FT-IR data will be discussed. Finally, all the VUV photoabsorption spectra in Figure~\ref{F6} show an underlying slope discussed in the next section.

\subsection{Rayleigh scattering from the ice residue}

\begin{figure}
\centering
%\hspace{-10.5mm}
\includegraphics[width=0.5\textwidth]{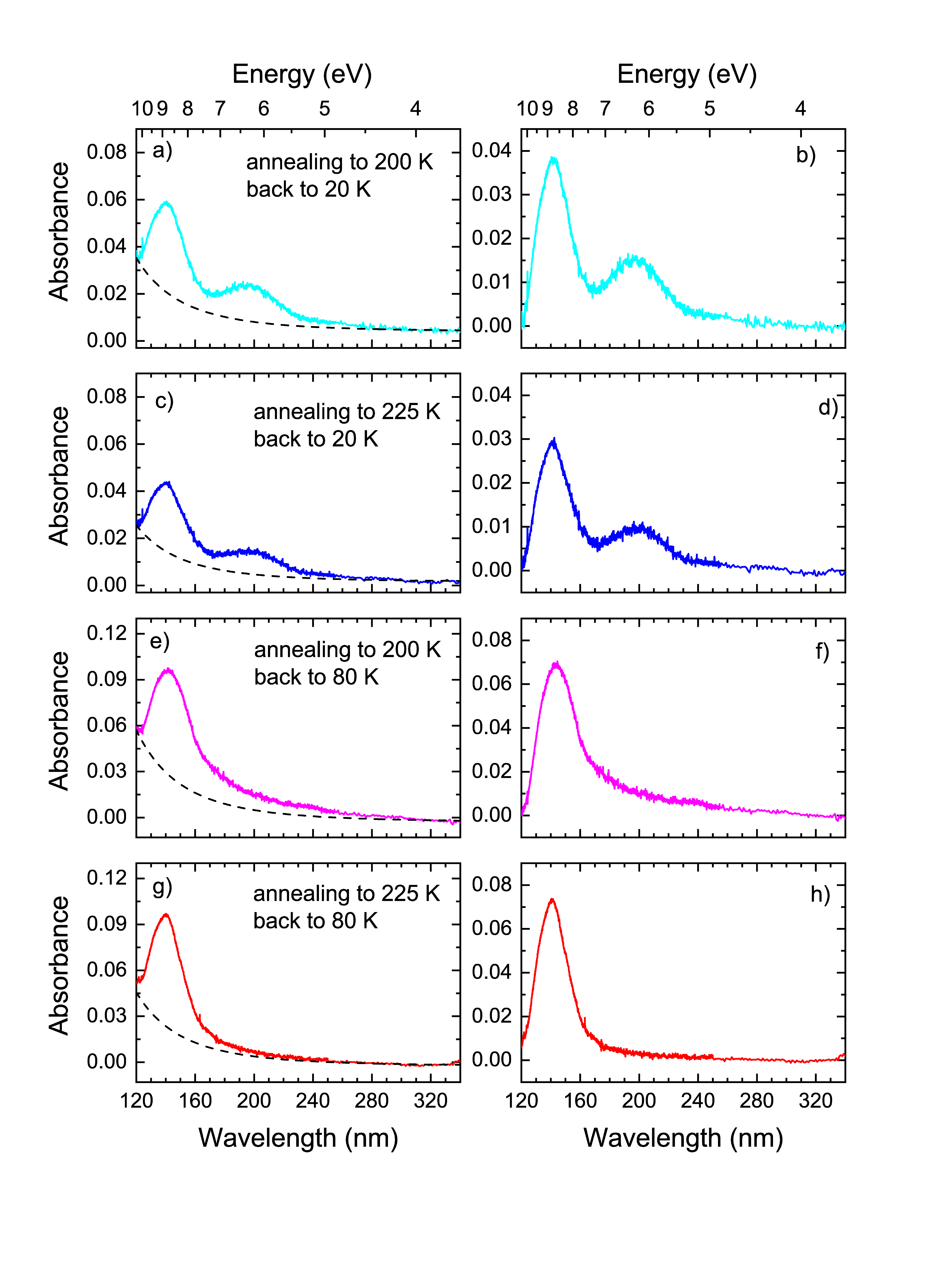}
\vspace{-5mm}
\caption{Left panels: VUV spectra of pure amorphous and crystalline H$_2$CO$_3$ ices at 20~K  and 80~K (solid lines) fitted with equation~\ref{slope} (dashed black lines) to highlight the effect of Rayleigh scattering. Right panels: VUV spectra of pure amorphous and crystalline H$_2$CO$_3$ ices at 20~K  and 80~K after fit subtraction to show real photoabsorption features of H$_2$CO$_3$ ice at 139 and 200~nm.}
\label{F7}
\end{figure}

The VUV photoabsorption spectra in Figure~\ref{F6} reveal an underlying slope that is potentially due to Rayleigh scattering off islands of material, that is H$_2$CO$_3$ ice, caused by a nonhomogeneous coverage of the substrate \citep{Dawes_etal2018}. Following \cite{Dawes_etal2018}, the scattering tails in the absorbance spectra of the H$_2$CO$_3$ ice can be fitted with a function of the form

\begin{equation}
		A~=~c\textrm{ln}\binom{1}{\overline{1-a\lambda^{-4}}} + b,
	\label{slope}
\end{equation}

\noindent
where the absorbance ($A$) equation, which is derived using the rearranged form of the Beer–Lambert Law, is modified by introducing the term $a\lambda^{-4}$ in the denominator corresponding to loss in the transmitted intensity due to scattering. Here $a$ is proportional to $r^6$, where $r$ is the scattering particle size and $c$ is proportional to the number density of the scatterers in the beam path. Figure~\ref{F7} shows the VUV photoabsorption spectra of H$_2$CO$_3$ ice together with the fitted curves (left panels) and the spectra after correction, that is subtracting the scattering tails (right panels). The good agreement in the spectral regions around 170 nm and $120-340$~nm between fits and spectra of H$_2$CO$_3$ synthesized from processing CO$_2$:H$_2$O~(6:1) ice mixtures at 20~K suggests that the observed slopes are likely due to Rayleigh scattering. On the other hand, the large discrepancies between fits and VUV spectra around the two absorption features at 139 and 200~nm, indicate that these are real absorption bands due to frozen H$_2$CO$_3$. In the case of solid H$_2$CO$_3$ synthesized from processed CO$_2$:H$_2$O~(6:1) ice mixtures at 80~K, the agreement between the fit and the spectra slope is at $180-340$~nm, indicating the presence of a single photoabsorption band at 139~nm. In conclusion, this section gives a further confirmation that both photoabsorption bands at 139 and 200~nm are real features of an H$_2$CO$_3$ ice, which is only a few nm thick and, therefore, does not fully and homogeneously cover the substrate surface causing Rayleigh light scattering in the VUV spectral range.

\subsection{Destruction of pure H$_2$CO$_3$ ice}

\begin{figure*}
\centering
%\hspace{-10.5mm}
\includegraphics[width=\textwidth]{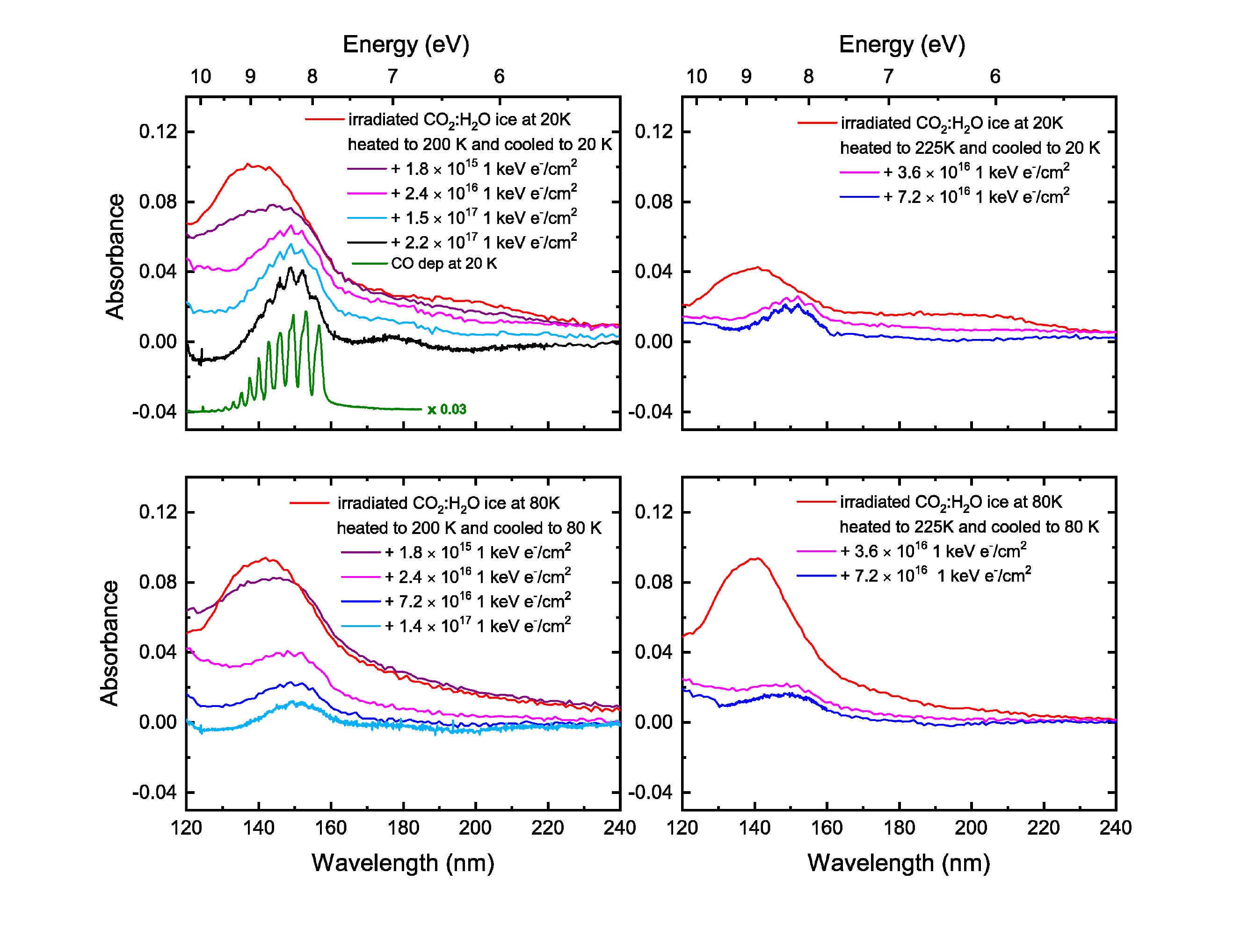}
\vspace{-5mm}
\caption{VUV ($120-240$~nm) photoabsorption spectra of pure amorphous and crystalline (left and right panels, respectively) H$_2$CO$_3$ ices (solid red lines) at 20 and 80~K (top and bottom panels, respectively) compared to the same ices after 1~keV electron irradiation at different fluences with maximum fluence of $2.2\times10^{17}$~e$^{-}$/cm$^{2}$. A VUV spectrum of pure CO ice is also displayed in top-left panel (solid green line).}
\label{F8}
\end{figure*}

\begin{table*}
\centering
	\caption{CO ice electronic transitions observed in the VUV photoabsorption spectra of 1~keV electron irradiated H$_2$CO$_3$ ice in the $120-160$~nm spectral range. Data are compared to the transition of pure CO ice deposited at 20~K.}
	\label{T3}
	\begin{tabular}{llllllll}
		\hline
        \hline
        \multicolumn{2}{c}{Pure CO}& \multicolumn{4}{c}{H$_2$CO$_3$ irr, 20~K}& \multicolumn{2}{c}{H$_2$CO$_3$ irr, 80~K}\\

        \multicolumn{2}{c}{}& \multicolumn{2}{c}{Ann, 200~K}& \multicolumn{2}{c}{Ann, 225~K}& \multicolumn{2}{c}{Ann, 200~K}\\

		[nm] & [eV] & [nm] & [eV] & [nm] & [eV]& [nm] & [eV]\\
		\hline
		127.1&  9.75&      &      &      &     &      &     \\
		129.0&  9.61&	   &      &      &     &      &     \\
        130.8&  9.48&	   &      &      &     &      &     \\
        133.0&  9.32& 132.4&  9.36&      &     &      &     \\
        135.3&  9.16& 134.8&  9.20&      &     &      &     \\
        137.6&  9.01& 137.1&  9.04& 137.9& 8.99&      &     \\
        140.1&  8.85& 140.5&  8.82& 140.2& 8.84&      &     \\
        142.8&  8.68& 143.0&  8.67& 142.5& 8.70& 143.0& 8.67\\
        145.6&  8.52&	   &      &      &     &      &     \\
        146.1&  8.49& 145.7&  8.51& 145.9& 8.50& 145.6& 8.52\\
        148.6&  8.34&	   &      &      &     &      &     \\
        149.6&  8.29& 148.8&  8.33&	148.6& 8.34& 149.2& 8.31\\
        152.0&  8.16&	   &      &      &     &      &     \\
        153.1&  8.10& 152.2&  8.15& 152.1& 8.15& 152.0& 8.16\\
        156.7&  7.91& 155.3&  7.98&		 &     &      &     \\
		\hline
	\end{tabular}
\end{table*}

\begin{figure*}
\centering
%\hspace{-10.5mm}
\includegraphics[width=\textwidth]{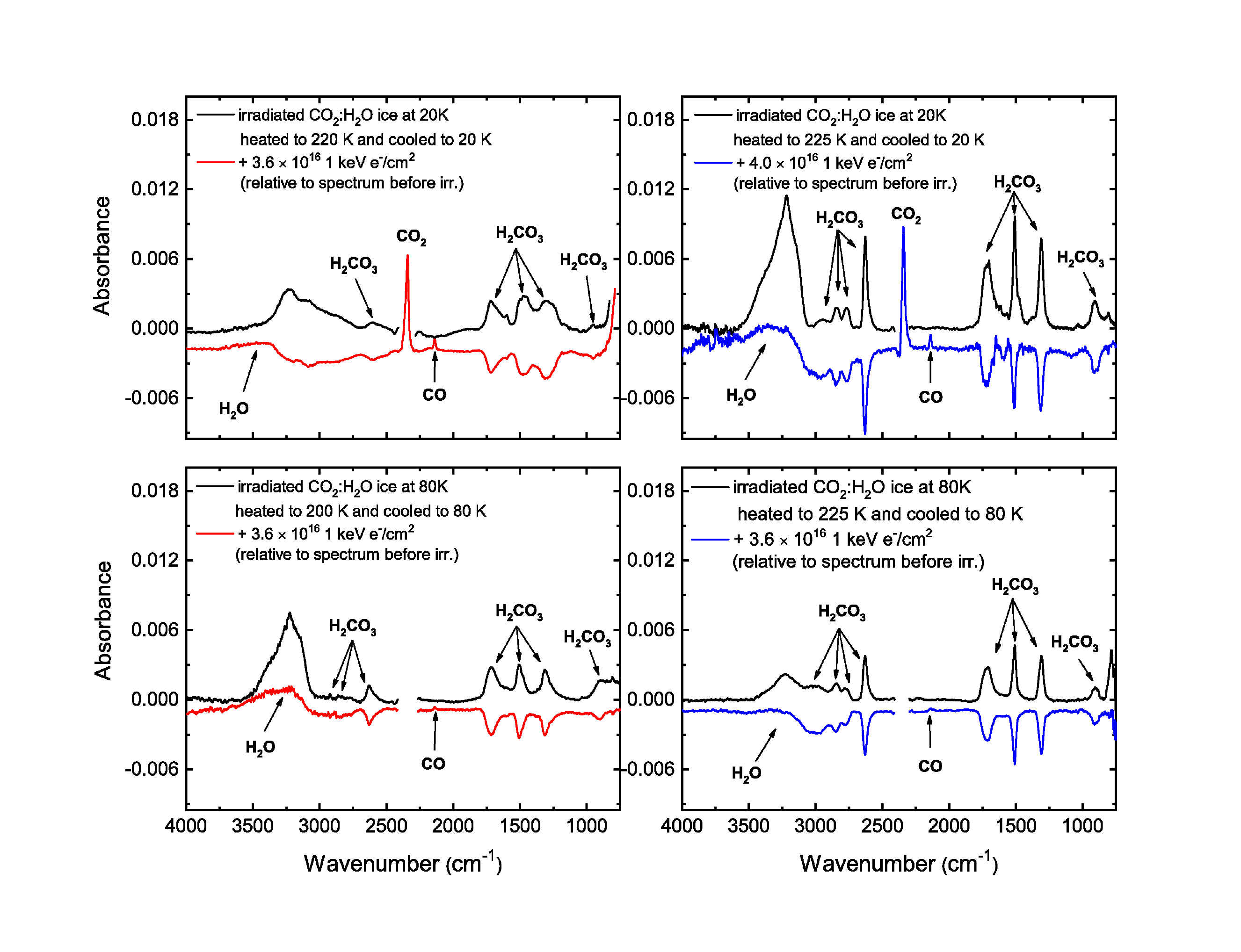}
\vspace{-5mm}
\caption{MIR ($4000-750$~cm$^{-1}$) FT-IR spectra of pure amorphous and crystalline (left and right panels, respectively) H$_2$CO$_3$ ices (solid black lines) at 20 and 80~K (top and bottom panels, respectively) compared to the same ices after 1~keV electron irradiation at different fluences with maximum fluence of $4.0\times10^{16}$~e$^{-}$/cm$^{2}$.}
\label{F9}
\end{figure*}

\begin{table*}
\centering
	\caption{MIR vibration mode peak positions of H$_2$CO$_3$ ice observed in FT-IR transmission spectra of 1~keV electron irradiated CO$_2$:H$_2$O~(6:1) ice mixtures annealed to different temperatures. Data are compared to literature values acquired in reflection mode at $10-20$~K.}
	\label{T4}
	\begin{tabular}{lllllll}
		\hline
        \hline
        \multicolumn{2}{c}{Literature}& \multicolumn{2}{c}{H$_2$CO$_3$ irr, 20~K}& \multicolumn{2}{c}{H$_2$CO$_3$ irr, 80~K}& Vibration\\

        $\beta$-H$_2$CO$_3$ & $\gamma$-H$_2$CO$_3$& Ann, 220~K& Ann, 225~K& Ann, 200~K &Ann, 225~K&\\

		[cm$^{-1}$]$^a$ & [cm$^{-1}$]$^b$ & [cm$^{-1}$] & [cm$^{-1}$] & [cm$^{-1}$] & [cm$^{-1}$]&\\
		\hline
		2850&  2853&  2892&  2850&  2855& 2841&O$-$H stretch\\
		2620&  2610&  2604&  2632&  2631& 2633&O$-$H stretch\\
        1723&  1766&  1722&  1726&  1714& 1715&C$=$O stretch\\
        1483&  1505&  1480&  1514&  1508& 1510&C$-$OH a-sym stretch\\
        1292&  1309&  1302&  1314&  1313& 1309&C$-$OH in-plane bend\\
        1038&  1030&  1026&  1037&  1040& 1036&C$-$OH sym stretch\\
            &   908&   903&   908&   900&  912&C$-$OH out-of-plane bend\\
         812&   809&      &   808&   801&  800&CO$_3$ out-of-plane bend\\
		\hline
	\end{tabular}
\begin{flushleft}
\footnotetext{}{$^{a}$\cite{Zheng_Kaiser2007} - CO$_2$:H$_2$O~(2.5:1) ice mixture exposed to 5~keV electrons at 10~K for 3 hours and then warmed to 210~K; $^b$\cite{Oba_etal2010} - Simultaneous deposition of CO and OH species on a 20~K gold surface followed by warm up to 220~K.}
\end{flushleft}
\end{table*}

The solid red lines in Figure~\ref{F8} are the non-normalized version of the VUV photoabsorption spectra of H$_2$CO$_3$ presented in Figure~\ref{F6} in the spectral range $120-240$~nm. It is interesting to report that the 139~nm absorption band intensity is comparable in all the spectra except for the crystalline H$_2$CO$_3$ at 20~K, which presents a much weaker feature. The 200~nm band intensity for both amorphous and crystalline H$_2$CO$_3$ is, however, comparable, with the one of crystalline ice being slightly less intense. All the other spectra shown in Figure~\ref{F8} were acquired after 1~keV electron irradiation of the residual H$_2$CO$_3$ at different fluences. \cite{Jones_etal2014b} studied 5~keV electron irradiation of carbonic acid at 80~K in the MIR showing the increasing appearance of water, CO$_2$, and perhaps traces of CO at different fluences. Our VUV irradiation spectra of solid H$_2$CO$_3$ showed a decrease of the H$_2$CO$_3$ absorption bands with increased fluence. The 200~nm band disappears almost immediately upon electron irradiation, indicating that it is caused by a metastable material such as $\gamma$-H$_2$CO$_3$ ice. The 139~nm band decreases more gently with increasing fluences, in line with the gradual formation of H$_2$O, CO$_2$, and CO ice. At fluences higher than $1\times10^{16}$~e$^{-}$/cm$^{2}$, the 139~nm band disappears leaving a new absorption feature at about 150~nm. This feature is consistent with a VUV absorption of a mixture of water and CO$_2$ ice, which reform upon electron processing. On top of the 150 nm feature the CO vibrational progression appears stronger with increasing fluences (see Table~\ref{T3} for peak positions). A closer comparison among the four experiments presented in Figure~\ref{F8} confirms that CO is more easily synthesized and preserved in the ice, when the latter is irradiated at lower temperature. This is also confirmed by the analog FT-IR data (see Figure~\ref{F9}). In general, lower yields are observed for both ices irradiated at 80~K when compared to those at 20~K and the smallest yields are seen in our experiments using crystalline H$_2$CO$_3$ ice irradiated at 80~K, where the CO vibrational progression is within the noise, that is below detection. At fluences higher than $1\times10^{17}$~e$^{-}$/cm$^{2}$, a new unknown feature appears at 180~nm. It could be an organic residue left on the surface of the substrate after the irradiation of the thin ice. However, more investigation is needed to provide any definitive characterization of this band.

Results shown in Figure~\ref{F9} are the analogs of those presented in Figure~\ref{F8}, but in the MIR range. Solid black lines are the H$_2$CO$_3$ ices as obtained from processing and annealing CO$_2$:H$_2$O~(6:1) ice mixtures. It should be mentioned that hexagonal ice (I$_\textrm{h}$) features around the 3~$\mu$m OH stretch water band are visible in some MIR spectra of annealed ice. Some water molecules are trapped in the H$_2$CO$_3$ ice independent of the conditions at which solid H$_2$CO$_3$ ice was formed, that are 20 or 80~K. The presence of a small amount of crystalline water in the H$_2$CO$_3$ does not, however, affect our results. Amorphous solid H$_2$CO$_3$ at 20~K presents broader and less intense MIR features if compared with crystalline solid H$_2$CO$_3$ at 20 K. New sharper peaks appear in the $3000-2500$ cm$^{-1}$ range in the crystalline ices. Those are due to the O-H stretch mode of H$_2$CO$_3$ ice \citep{Zheng_Kaiser2007}. Table~\ref{T4} compares the MIR vibrational modes of H$_2$CO$_3$ ice obtained in our experiments with literature data of $\beta$- and $\gamma$-H$_2$CO$_3$ ice \citep{Zheng_Kaiser2007, Oba_etal2010}. The profile of amorphous H$_2$CO$_3$ ice obtained after electron irradiation of a CO$_2$:H$_2$O~(6:1) ice mixtures at 20~K followed by annealing to 220~K is qualitatively closer to the one from \cite{Oba_etal2010}, that is $\gamma$-polymorph ice. Absorption band peak positions are also shifted compared to the other H$_2$CO$_3$ ice (see Table~\ref{T4}). This is particularly evident from the weak absorption band around 2600~cm$^{-1}$, that is the O$-$H stretch mode. In the case of amorphous H$_2$CO$_3$ ice at 80~K, the band is sharper and shifted by some 30~cm$^{-1}$; it becomes even more intense for crystalline H$_2$CO$_3$ ices. FT-IR spectra of amorphous and crystalline H$_2$CO$_3$ at 80~K are qualitatively similar to literature spectra of $\beta$-polymorph ice \citep{Gerakines_etal2000, Zheng_Kaiser2007}. This is in good agreement with our assignment of the $\beta$- and $\gamma$-H$_2$CO$_3$ ices in our VUV photoabsorption data. Hence, MIR spectra seems to be more sensitive to amorphous-to-crystalline transitions and less sensitive to the H$_2$CO$_3$ polymorphism than VUV data.

The solid red and blue lines in Figure~\ref{F9} are FT-IR spectra of 1~keV electron irradiated amorphous and crystalline H$_2$CO$_3$ ice, respectively. We note that infrared spectra of H$_2$CO$_3$ before irradiation were used as reference spectra, hence, the negative peaks mirroring those from solid H$_2$CO$_3$. Therefore, negative peaks show the destruction of H$_2$CO$_3$ ice, while positive peaks the formation of new species in the ice upon electron exposure. Water, CO, and CO$_2$ are confirmed to be formed from irradiation of H$_2$CO$_3$ ice. As discussed before, CO is best synthesized and retained in the ice irradiated at 20~K. Only traces of CO are visible in the experiments at 80~K, in agreement with \cite{Jones_etal2014a}. Unfortunately, CO$_2$ is only visible in our FT-IR data, when gas-phase absorptions from CO$_2$ along the IR-beam line-of-sight outside the vacuum chamber were properly subtracted by the reference spectra. This was the case only for ices processed at 20~K, hence, the decision to mask the CO$_2$ stretch region for the other spectra. In the experiments at 20~K, a larger amount of solid CO$_2$ is formed in the crystalline H$_2$CO$_3$ ice.

\cite{Jones_etal2014a} showed that CO$_2$ formed at 80~K by 5~keV electron irradiation of crystalline H$_2$CO$_3$ ice (i.e., ice annealed at 220~K) present peculiar band profiles and peak positions that do not resemble CO$_2$ synthesized from other ice species containing carbon and oxygen. The authors suggested that CO$_2$ from irradiation of H$_2$CO$_3$ ice fit well an absorption feature observed on Callisto. The \cite{Jones_etal2014a} experiment is analog to our experiment shown in the bottom-right panel of Figure~\ref{F9}. Unfortunately, we are not able to check the newly synthesized CO$_2$ band profile and peak position for the 80~K experiment because of the aforementioned issue with gas-phase CO$_2$ contribution outside the chamber. However, regarding CO$_2$ synthesized after irradiation at 20~K of amorphous and crystalline H$_2$CO$_3$ ice, the peak positions for both are at 2343~cm$^{-1}$ and the band is asymmetric with a profile similar to CO$_2$ mixed with H$_2$O and a small amount of CO ice at 20~K \citep{Whittet_etal1998}. Thus, similar experiments carried-out at different temperatures lead to the formation of the same species, for example CO$_2$, but with absorption band profiles substantially different from each other and temperature dependent.

\section{Astrophysical implications}

The systematic laboratory work presented here on the formation and destruction of amorphous and crystalline carbonic acid ice has implications for a large range of astrophysical environments. Although H$_2$CO$_3$ is not among the around 200 molecules currently detected in the interstellar medium or circumstellar shells, there is laboratory evidence that H$_2$CO$_3$ is embedded in a water-rich layer on interstellar grains. \cite{Oba_etal2010} studied the H$_2$CO$_3$ formation upon the surface reaction route starting from CO~+~OH to form $cis$- and $trans$-HOCO followed by the HOCO~+~OH reaction to form H$_2$CO$_3$. This surface route can potentially be important for the formation of carbonic acid in cold dense interstellar clouds, where atom addition reactions are dominant on ice grains. In dense clouds, only cosmic rays and the cosmic ray-induced UV field can further process ice mantles. On the other hand, energetic bombardment (e.g., ions, photons, and electrons) of CO$_2$:H$_2$O mixtures in a wide range of ratios also leads to the formation of solid H$_2$CO$_3$ \citep{Moore_etal1991a, Gerakines_etal2000, Zheng_Kaiser2007}. H$_2$O and CO$_2$ are among the most abundant species found in the solid phase in space and are also found mixed within the same interstellar ice layer \citep{Boogert_etal2015}. Hence, our laboratory work suggest that a combination of $\beta$- and $\gamma$-H$_2$CO$_3$ ice should be searched for in the ISM, with a prevalence of $\gamma$-H$_2$CO$_3$ ice in cold dark clouds and $\beta$-H$_2$CO$_3$ ice in more evolved star-formation regions, such as high- and low-mass young stellar objects (YSOs). A possible explanation for the non-detection of gas-phase H$_2$CO$_3$ in the ISM is its extremely low vapor pressure under vacuum. H$_2$CO$_3$ is a refractory species and can stay on interstellar grains beyond water sublimation. Thus, H$_2$CO$_3$ has a chance to be included in protoplanetary accretion disks, planets, moons and comets without going from the solid to the gas phase. On the other hand, we show that H$_2$CO$_3$ ice can easily be destroyed by electron bombardment of the ice. Hence, if H$_2$CO$_3$ is not shielded from energetic triggers, it will dissociate into CO, CO$_2$, and H$_2$O ice. Another issue linked to the detection of H$_2$CO$_3$ on interstellar grains is its spectral confusion in the MIR range. Some of the weak ice absorption features in the $3.3-4.0$, $7.0-7.9$, and $8.0-10.4$~$\mu$m ($3030-2500$, $1428-1265$, and $1250-961$~cm$^{-1}$) spectral regions observed toward YSOs could be at least partially due to solid H$_2$CO$_3$. Future space-based IR facilities such as the JWST with high spectral resolving powers $R$ of up to 3,000 are set to provide strong constraints on the identification of new ice species in the aforementioned IR spectral ranges. Our systematic work on solid H$_2$CO$_3$ at different temperatures and morphologies shows that MIR band profiles for amorphous H$_2$CO$_3$ ice are broader and weaker than the corresponding bands of crystalline H$_2$CO$_3$ ice. This is also the case for $\gamma$-H$_2$CO$_3$ compared to $\beta$-H$_2$CO$_3$ ice. Although the $\beta$-polymorph seems to be the most stable form of H$_2$CO$_3$ ice, a $\gamma$-H$_2$CO$_3$ structure of the ice is potentially present in space and should be searched for in the coldest regions of the ISM and Solar System by JWST.

Frozen CO$_2$ and H$_2$O have been detected on the surface of many objects exposed to energetic processing in the inner and outer Solar System, for example planet polar caps as well as ice surfaces of moons, comets, and distant small Solar System bodies. For instance, Mercury's north pole and Martian polar caps are continuously bombarded by solar wind particles. Most of the icy moons belonging to Jupiter and Saturn are embedded in the magnetospheres of their respective planets, hence, exposed to energetic ions and electrons. Finally, comets, trans-Neptunian objects (TNOs) and Kuiper belt objects (KBOs) are all exposed to galactic cosmic rays and solar wind ion bombardment for billions of years. Thus, it is likely that solid H$_2$CO$_3$ is synthesized and destroyed on the surface of the aforementioned objects. Our set of VUV data indicate that solid H$_2$CO$_3$ presents two distinct photoaborption bands in the $120-340$ nm range at 139 and 200~nm that we have assigned to $\beta$- and $\gamma$-H$_2$CO$_3$ ice, respectively (see Figure~\ref{F7}). Although $\beta$-polymorph ice is synthesized upon any energetic processing, $\gamma$-H$_2$CO$_3$ ice forms in presence of abundant CO ice at low temperatures. Therefore, we expect $\beta$-polymorph ice to be ubiquitously formed in the Solar System and $\gamma$-polymorph to be present mainly in the outer Solar System beyond the CO snowline. Our work shows that VUV spectroscopy can potentially be a key tool in the identification of different forms of H$_2$CO$_3$ ice.

\begin{figure}
\centering
%\hspace{-10.5mm}
\includegraphics[width=0.5\textwidth]{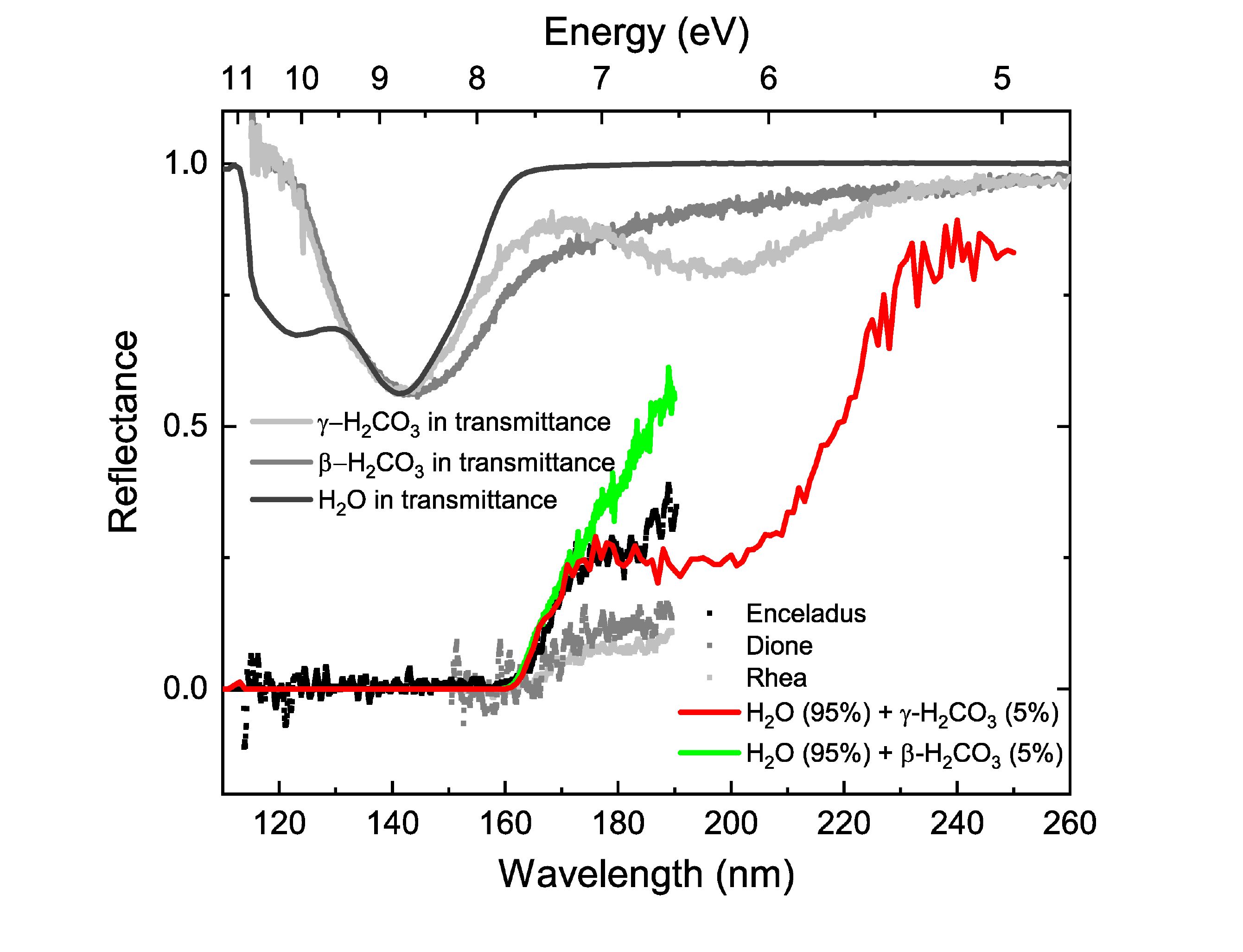}
\vspace{-5mm}
\caption{VUV ($115-190$~nm) Cassini UVIS data of the surface of Enceladus (square black symbols), Dione (square dark gray symbols), and Rhea (square light gray symbols) compared to two fits: a fit (solid red line) of the VUV photoabsorption spectra of pure ASW (solid black line) deposited at $\sim$20~K and amorphous $\gamma$-H$_2$CO$_3$ at 20~K (solid light gray line); and a fit (solid green line) of the VUV photoabsorption spectra of the same water ice and amorphous $\beta$-H$_2$CO$_3$ at 80~K (solid dark gray line). The laboratory ice components are normalized and scaled in the figure for clarity.}
\label{F10}
\end{figure}

In Figure~11 of \cite{Ioppolo_etal2020}, we compared the reflectance spectra of Enceladus, Dione, and Rhea, as measured in the VUV ($115-190$~nm) by the Cassini UltraViolet Imaging Spectrograph \citep[UVIS, ][]{Hendrix_etal2010}, with a two-components fit of the VUV transmittance spectra of ASW and nitrogen dioxide (NO$_2$), both deposited at $\sim$20~K. As discussed in \cite{Ioppolo_etal2020}, the observational spectra of the Saturn's moons present a saturated band between $115-160$~nm that is reproduced by a VUV spectrum of ASW and a slope between $160-200$ nm, peaking at 180~nm, that is fitted with a VUV spectrum of solid NO$_2$. In Figure~\ref{F10}, we show the fit of the observational data of Enceladus with VUV transmittance spectra of pure ASW at $\sim$20~K and amorphous carbonic acid to test whether $\beta$- and/or $\gamma$-H$_2$CO$_3$ are potentially present on the surface of the moons of Saturn. The figure shows that a VUV spectral fit of 95\% ASW at $\sim$20~K and 5\% amorphous $\gamma$-H$_2$CO$_3$ at 20~K (solid red line) cannot reproduce observations between $180-200$~nm because the 200~nm absorption band of $\gamma$-H$_2$CO$_3$ appears to be too broad in the $190-230$~nm region. The far ultraviolet (FUV) spectrum of Saturn’s moon Enceladus from the Cosmic Origins Spectrograph (COS) on the Hubble Space Telescope (HST, not shown in Figure~\ref{F10}) combined with Cassini UVIS data, as shown in Figure~6 of \cite{Zastrow_etal2012}, reveals that the VUV spectrum of Enceladus does not have any clear absorption feature in the $200-260$~nm region. Moreover, Figure~3 of \cite{Hendrix_etal2018} compares the VUV-UV-vis composite spectra of several Saturn's moons including Enceladus, Dione, and Rhea. In all cases, there is not a clear absorption feature centered at 200~nm, while the absorption at $\sim$250~nm has been already assigned to solid ozone \citep[e.g., see][]{Noll_etal1997, Boduch_etal2016}. Therefore, although traces of $\gamma$-H$_2$CO$_3$ on the surface of Saturn's satellites cannot be completely ruled-out, $\gamma$-H$_2$CO$_3$ should not be considered as a major fit component of the VUV observational spectra.

A VUV spectral fit of 95\% ASW at $\sim$20~K and 5\% amorphous $\beta$-H$_2$CO$_3$ at 80~K (solid green line in Figure~\ref{F10}) can reproduce VUV observations of Enceladus between $115-170$~nm. In this case, the observed 180 nm absorption slope is not well fitted. However, the 180~nm feature can be reproduced by a third component of NO$_2$ (not shown in the figure) as suggested in \cite{Ioppolo_etal2020}. Hence, our results are still in agreement with those in \cite{Ioppolo_etal2020}. The addition of $\beta$-H$_2$CO$_3$ as a VUV spectral component to the fit shown in \cite{Ioppolo_etal2020} causes a slight reduction in the amount of ASW needed for the fit, while improving it in the $160-170$~nm region. It should be, however, noted that although the mean surface temperatures of Endeladus, Dione, and Rhea around $70-90$~K are consistent with the formation of $\beta$-H$_2$CO$_3$ in H$_2$O:CO$_2$ mixtures, both the 139~nm $\beta$-H$_2$CO$_3$ and 144~nm ASW features lie in the saturated region of the observational spectra. Hence, a quantitative analysis of the fit components as well as the unambiguous identification of carbonic acid on the surface of Saturn's moons are not possible at this stage. Further observational evidence of solid carbonic acid in the Solar System is expected to be delivered by the JUICE mission. Unfortunately, the instruments on board JUICE spacecraft will not cover the $210-400$~nm spectral range. Hence, an unambiguous detection of the 200~nm band of $\gamma$-H$_2$CO$_3$ may prove to be challenging. However, for all the aforementioned reasons, the 139~nm band of both $\beta$- and $\gamma$-H$_2$CO$_3$ is a good candidate for the unambiguous identification of solid H$_2$CO$_3$ on the $50-100$~K ice surfaces of Jupiter's moons by the upcoming JUICE mission, provided that the VUV spectra of the Jovian moons will not be saturated in the $100-200$~nm spectral range.

\section{Conclusions}
We have presented new VUV absorption spectra of pure and mixed CO$_2$ and H$_2$O ices exposed to 1\,keV electron irradiation under conditions relevant to the ISM and the Solar System. Below we list the main findings of this work.

The 1\,keV electron irradiation of solid pure CO$_2$ ice at 20~K leads to the formation of frozen CO and O$_3$ as detected by their VUV photoabsorption features. Products of the electron irradiation of CO$_2$ ice at 75~K desorb efficiently upon formation, leaving only traces of trapped CO in the ice, while O$_3$ is under our detection limit. VUV spectra of pure water ice before and after electron irradiation at 20~K are similar to each other, indicating that the electron dissociated products are likely to recombine in the ice. At 80~K, desorption of the ice is indirectly monitored by the loss of ice due to the sublimation of volatile reaction products in the upper layers of the ice.

The 1\,keV electron irradiation of CO$_2$:H$_2$O~(6:1) ice mixtures at 20 and 80~K shows the formation of CO and O$_3$ molecules in the VUV spectral range, with the electron irradiation products being more abundantly present at low temperature. MIR spectra of the corresponding ices reveal the formation of H$_2$CO$_3$ both at 20 and 80~K and the trapping of its precursors, that are the $cis$- and $trans$-HOCO radicals, at 20~K.

We have shown the first systematic VUV spectra of pure amorphous and crystalline H$_2$CO$_3$ ice obtained by annealing the CO$_2$:H$_2$O~(6:1) ice to 200 and 225~K, respectively, after electron irradiation at 20 and 80~K. Two photoabsorption bands are assigned to solid H$_2$CO$_3$ at 139 and 200~nm with the latter being only detected when H$_2$CO$_3$ was synthesized at 20~K. Hence, we have assigned the 139~nm band to $\beta$-H$_2$CO$_3$, the form of H$_2$CO$_3$ notoriously produced by energetic processing of CO$_2$- and H$_2$O-rich ices; and we have assigned the 200~nm band to a different form of H$_2$CO$_3$, the so-called $\gamma$-polymorph ice, produced through the surface reactions between CO molecules and OH radicals occurring at low temperatures, that is below 30~K. Assignments are also supported by complementary MIR data.

We provide a complete set of VUV spectra at different temperatures in support of the future identification of H$_2$CO$_3$ ice in space. A qualitative comparison of our laboratory VUV spectra with Cassini UVIS data of the ice surface of Enceladus, Dione, and Rhea is discussed. Our data indicate that $\beta$-H$_2$CO$_3$ should be ubiquitously present in space on the surface of CO$_2$- and H$_2$O-rich ices, while unique spectral features of $\gamma$-H$_2$CO$_3$ should be searched for in the coldest regions of the ISM and in the outer Solar System beyond the CO snowline.

\begin{acknowledgement}
The authors thank Dr. Amanda R. Hendrix for providing observational Cassini's UVIS data and the anonymous reviewers whose suggestions helped improve and clarify this manuscript. The research presented in this work has been supported by the Royal Society University Research Fellowship (UF130409), the Royal Society University Research Fellowships Renewal (URF/R/191018), the Royal Society Research Fellow Enhancement Award (RGF/EA/180306), and the project CALIPSOplus under the Grant Agreement 730872 from the EU Framework Programme for Research and Innovation HORIZON 2020. Furthermore, S.I. recognizes the Royal Society for financial support. The research of Z.K. is supported by VEGA – the Slovak Grant Agency for Science (grant No. 2/0023/18) and COST Action TD 1308. R.L.J. acknowledges the STFC for her Ph.D. Studentship under grant no. ST/N50421X/1. A.D. acknowledges the Daphne Jackson Trust and Leverhulme Trust (grant no. ECF/2016-842) for her Fellowships. R.L.J. and A.D. also acknowledge the Open University for financial support. N.J.M. acknowledges support from the Europlanet 2020 RI and the Europlanet 2024 RI, which have received funding from the European Union's Horizon 2020 research and innovation programme under grant agreements No 654208 and 871149, respectively. G.S. is supported by the Italian Space Agency (ASI 2013-056 JUICE Partecipazione Italiana alla fase A/B1) and by the European COST Action TD1308-Origins and evolution of life on Earth and in the Universe (ORIGINS).
\end{acknowledgement}

\begin{appendix}
\section{1~keV electron irradiation of pure CO$_2$ and H$_2$O ices at $\sim$80~K}

\begin{figure}
\centering
\includegraphics[width=0.5\textwidth]{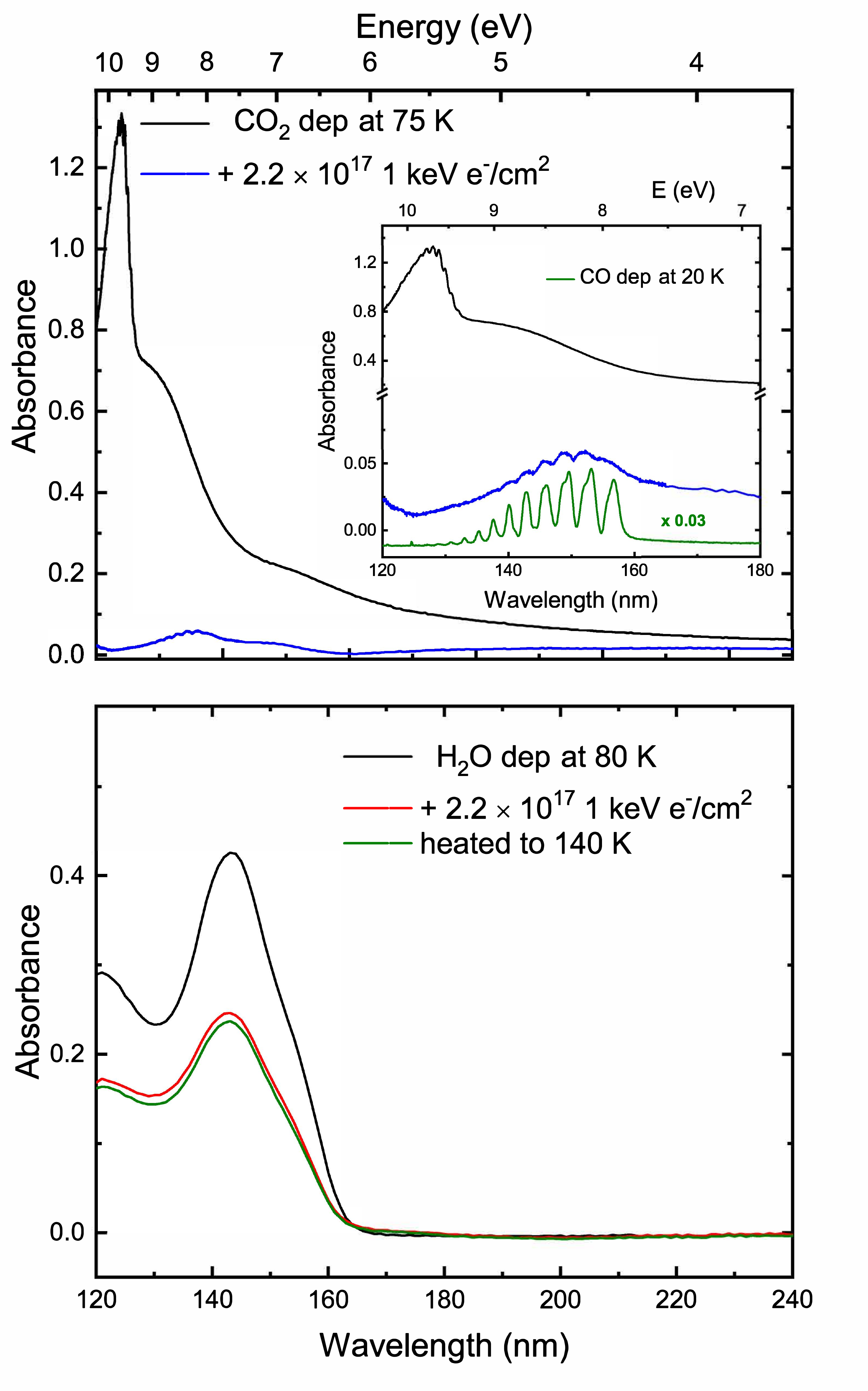}
\vspace{-5mm}
\caption{Top panel: VUV ($120-340$~nm) photoabsorption spectra of pure CO$_2$ ices deposited at 75~K before and after (solid black and blue lines, respectively) 1~keV electron irradiation with a total fluence of $2.2\times10^{17}$~e$^{-}$/cm$^{2}$ (i.e., a dose of $9.9\times10^{2}$~eV/16u). A VUV spectrum of pure CO ice is also displayed (solid green line). Bottom panel: VUV ($120-340$~nm) photoabsorption spectra of pure H$_2$O ices deposited at 80~K before and after (solid black and red lines, respectively) 1~keV electron irradiation with a maximum total fluence as for pure CO$_2$ ice. The VUV photoabsorption spectrum of the processed ice heated to 140~K is also shown (solid green line).}
\label{S1}
\end{figure}

The Top panel of Figure~\ref{S1} displays VUV photoabsorption spectra of crystalline CO$_2$ ice deposited at 75~K and bombarded with 1~keV electrons at the same temperature. The VUV spectrum of deposited ice (solid black line) is slightly different from the corresponding one deposited at 20~K (see Figure~\ref{F2}). The peak position of the 126~nm band is shifted to longer wavelengths and the profile appears broader with a less prominent vibrational progression (see inset of top panel of Figure~\ref{S1}). It should be noted that the 126~nm spectral band of the ice deposited at 75~K presents a relative absorbance larger than 1 at its peak. Therefore, some of the differences between the 126 nm band profiles at 20 and 75~K can also be attributed to saturation effects in the spectrum at 75~K. The 140~nm band is also broader, and appears as a shoulder on the larger 126~nm band. The VUV spectrum of the irradiated ice (solid blue line) shows that basically all CO$_2$ molecules desorbed during electron bombardment. This is not surprising because pure CO$_2$ molecules desorb around 80~K \citep{Collings_etal2004}. A small amount of CO and CO$_2$ is trapped in the ice residue, which is likely made of the background water sputtered from the chamber walls during irradiation. Peak positions of the CO vibrational progression are shifted compared to those of a pure CO ice (see solid green line in the inset of the top panel of Figure~\ref{S1}). It is also likely that species formed in the ice, like CO and O$_3$, co-desorbed with CO$_2$ quickly after their formation because of the high temperature of the sample.

The bottom panel of Figure~\ref{S1} shows the VUV photoabsorption spectra of low density ASW deposited (solid black line) and irradiated (solid red line) at 80~K and then heated to 140~K (solid green line). The 144~nm band of the VUV spectrum of ASW deposited at 80~K presents a more defined shoulder at 155~nm and a more pronounced local minimum at 132~nm compared to the ones of the high density ASW sample deposited at 20~K as shown in Figure~\ref{F3}. It should be highlighted that during deposition of ASW at 80~K, the HeNe laser signal was lost and the ice thickness was not measured. As stated in the main text, deposition was controlled by keeping the deposition gas pressure inside the main chamber and the time of deposition the same as for the deposition of ASW at 20~K. A final roughly $2-3$ times thinner ice was deposited at 80~K compared to the deposition at 20~K according to the photoabsorption relative intensities of the 144~nm band shown in Figures~\ref{F3} and \ref{S1}. Some desorption was indirectly observed upon 1~keV electron irradiation of the ice at 80~K. The different temperature and thickness, and unknown deposition rate of this ice due to technical issues during the beamtime at ASTRID2 can explain the observed relatively high loss of water ice upon 1~keV electron processing compared to the experiment at 20~K. All VUV photoabsorption bands of ASW before and after irradiation at 80~K are qualitatively similar to each other suggesting that no other species is formed in the ice as for the ASW experiment at 20~K. Finally, the VUV spectrum of the ice heated to 140~K slightly deviate from the irradiated spectrum at 80~K in agreement with our VUV analog data at 20~K.

\end{appendix}

\end{document}